\documentclass[11pt]{article}

\usepackage{amssymb,amsmath, epsfig}
\usepackage{changepage}
\usepackage{caption}
\usepackage[section]{placeins}

\def\u{\vskip  .075 in}
\def\nh{\noindent\hangindent=1 true cm \hangafter = 1}

\def\nh{\noindent\hangindent=1 true cm \hangafter = 1}

\def\u{\vskip  .1 in}

\def\B {\begin{eqnarray*}}

\newcommand{\bel}[1]{\begin{equation}\label{#1}}

\newcommand{\be}{\begin{equation}}
\newcommand{\qe}{\end{equation}}
\newcommand{\ee}{\end{equation}}
\newcommand{\baS}{\begin{eqnarray}}
\newcommand{\ba}{\begin{eqnarray}}
\newcommand{\ea}{\end{eqnarray}}

\def\CA{${\rm Ca^{2+}}$ }

\def\EN{\end{eqnarray*}}

\def\ac{${\rm Ca_v1.3}$ }

\begin{document}
\title{Analysis of pacemaker activity in a two-component model of 
some brainstem neurons\\
\   \\
{\normalsize
Henry C. Tuckwell$^{1,2*}$, Ying Zhou$^{3,}$, Nicholas  J. Penington $^{4,5}$\\   \
\  \\ 
 \              \\
$^1$ School of Electrical and Electronic Engineering, University of Adelaide,\\
Adelaide, South Australia 5005, Australia \\
$^2$ School of Mathematical Sciences, Monash University, Clayton, Victoria 3800, Australia\\
\   \\
$3$ Department of Mathematics, Lafayette College, 1 Pardee Drive, Easton, PA 18042, USA\\
\              \\
$^4$ Department of Physiology and Pharmacology,\\
$^5$ Program in Neural and Behavioral Science and Robert F. Furchgott
Center for Neural and Behavioral Science \\
State University of New York,
Downstate Medical Center,\\
Box 29, 450 Clarkson Avenue, Brooklyn, NY 11203-2098, USA\\
\     \\
$^{*}$ {Corresponding author:} Henry C. Tuckwell, School of Electrical and Electronic Engineering, University of Adelaide, North Terrace, Adelaide, SA 5005,
Australia; \ \\
Tel: 61-481192816; Fax: 61-8-83134360: email:   
 henry.tuckwell@adelaide.edu.au \\
\     \\
 }}

\maketitle

\newpage 
\begin{abstract}  
Serotonergic, noradrenergic and dopaminergic brainstem (including midbrain) neurons, 
often exhibit spontaneous  and fairly regular spiking with frequencies of order 
a few Hz, though dopaminergic and noradrenergic neurons only exhibit such pacemaker-type activity in vitro or in vivo under special conditions.
  A large number of ion channel types contribute to such spiking so that
detailed modeling of spike generation leads to the requirement of solving very large
systems of differential equations. It is useful to have simplified mathematical models 
of spiking in such neurons so that, for example, features of
inputs and output spike trains can be incorporated including stochastic effects for possible use in network models.   
 In this article we investigate
a simple two-component conductance-based model of the Hodgkin-Huxley type. Solutions are computed numerically and  with suitably chosen parameters mimic features of pacemaker-type spiking in
the above types of neurons.  The effects of varying parameters is investigated 
in detail, it being found that there is extreme sensitivity to eight of them. 
Transitions from non-spiking to spiking are examined for two
of these, the half-activation potential for an activation variable and the
added (depolarizing) current and contrasted with the behavior of the
classical Hodgkin-Huxley system. 
The plateaux levels between spikes can be adjusted, by changing
a set of voltage parameters, to agree with experimental observations.
Experiment has shown that in, in vivo, dopaminergic and noradrenergic neurons' pacemaker activity can be  induced by the removal of excitatory inputs or the introduction of inhibitory ones. These properties are confirmed by mimicking opposite such changes in the model, which resulted in
a change from pacemaker activity to bursting-type phenomena.  The article concludes with a brief review of previous modeling of these types of neurons and a brief discussion of  their involvement in some pathological pysychiatric conditions.
\end{abstract}


\rule{80mm}{.5pt}
\tableofcontents
\rule{80mm}{.5pt}

\noindent {\bf Abbreviations}  \\
\noindent 5-HT, 5-hydroxytryptamine (serotonin); AC, adenylate cyclase; 
AHP, afterhyperpolarization; cAMP, cyclic adenosine monophosphate;
CREB, cAMP response element binding protein; CRH, corticotropin-releasing hormone; CRN, caudal raphe
nucleus; DA, dopamine or dopaminergic; 
 DRN, dorsal raphe nucleus;  
 GABA, gamma-aminobutyric acid; HH, Hodgkin-Huxley; Hz, hertz; HPA, hypothalamus-pituitary-adrenal;  ISI, interspike interval; LC, locus coeruleus; MDD, major depressive disorder; NA, noradrenaline or noradrenergic; OCD, obsessive-compulsive disorder; PTSD, post-traumatic stress disorder; PVN, paraventricular nucleus (of hypothalamus); REM, rapid eye movement; SE, serotonin or serotonergic.


%

  \section{Introduction}
Neurons which exhibit (approximately) periodic spiking in the presumed absence of synaptic input are called  autonomous pacemakers, and include neurons found 
in the subthalamic nucleus, nucleus basalis, globus pallidus, raphe nuclei,
cerebellum, locus ceruleus, ventral tegmental area, and substantia nigra (Ramirez et al., 2011). In order to sustain spiking, some pacemaking cells may require small amounts of depolarizing
inputs, natural or laboratory. Thus, for example Pan et al. (1994) found that all of 42
rat LC neurons fired spontaneously, whereas in Williams et al. (1982) and 
Ishimatsu et al. (1996) it was reported that most cells did not require excitatory input to
fire regularly.  These latter three sets of results were obtained in vitro.
Sanchez-Padilla et al. (2014) reported that spike rate in mouse LC neurons  
was not affected by blockers of glutamatergic or GABAergic synaptic input,
supporting the idea that these cells were autonomous pacemakers. 
In Grace and Onn (1989), it was found that in vitro, rat midbrain dopamine neurons fired periodically with a frequency which could be increased or decreased by the application of depolarizing or hyperpolarizing currents without changing the pattern of spikes
- see Figure 1 herein. The mechanisms of pacemaking activity of DA neurons in the rat
ventral tegmental area (slice) was analyzed in Khaliq and Bean (2008).

The neurons with which we are mainly concerned  are serotonergic neurons
of the (dorsal) raphe nucleus, noradrenergic neurons of the locus coeruleus and midbrain dopaminergic neurons.   The electrophysiological properties of
these cells have been much investigated over the last several decades (Aghajanian and Vandermaelen, 1982; Aghajanian et al., 1983;
 Alreja and Aghajanian, 1991; Williams et al., 1984; Grace and Bunney, 1983, 1984; Harris et al., 1989). 
The roles of serotonergic and noradrenergic neurons in stress related
disorders such as MDD and PTSD  by means of reciprocal  interactions with, inter alia, 
the HPA axis (especially through the PVN),  hippocampus and prefrontal cortex are well documented (Lopez et al., 1999; Carrasco and Van de Kar, 2003;
Lanfumey et al., 2008; Mahar et al., 2014). For example, CRH neurons
of the PVN project directly to SE neurons of the DRN (Lowry et al., 2000) and
NA neurons of the LC (Valentino et al., 1993) and 
stress, via an upregulation of the cAMP pathway,  
leads to an increase in the excitability of LC neurons (Nestler et al., 1999).  In addition many of these
brainstem neurons are endowed with glucocorticoid receptors 
which are activated by high levels of corticosterone (cortisol) (Jo\"els et al., 2007). 

 In most common experimentally employed animals except cat, the locus coeruleus is almost completely
homogeneous, consisting of noradrenergic neurons which in rat number about
1500 (Swanson, 1976; Berridge and Waterhouse, 2003). 
 The number of neurons in the rat DRN is between about 12000 and 15000
(Jacobs and Azmitia, 1992; Vertes and Crane, 1997) of which up to 
50\% are principal serotonergic cells (Vasudeva and Waterhouse, 2014)  but there are also present ~1000 dopaminergic cells (Lowry et al., 2008) and 
 GABAergic cells, whose density varies throughout the divisions of the nucleus,  as well as several other types of neuron. In rat, the number of dopamine neurons (bilateral counts) in the midbrain
groups is approximately 71,000, with 40,000 in the VTA and 
 25,000 in substantia nigra (German and Manaye, 1993; 
Nair-Roberts et al., 2008). In the latter two cell groups there
are also about 21,000 and 41,000 GABA-ergic cells, respectively.

 The SE neurons of the 
 DRN and NA neurons of the LC,  often exhibit a slow regular pattern
of firing with frequencies of order 0.5 to 2 Hz in slice (Milnar et al., 2016)  and sometimes higher
in vivo.  Midbrain dopaminergic neurons exhibit similar firing patterns in vitro. The origins of pacemaker firing differ amongst 
various neuronal types.
Thus, brainstem dopaminergic neurons may  
fire regularly without  excitatory synaptic input (Grace and Bunney
1983; Harris et al., 1989). Underlying the rhythmic activity are subthreshold oscillations
which were demonstrated with a mathematical model to possibly reflect an interplay between an L-type calcium current and a calcium-activated
potassium current (Amini et al, 1999) - but see Kuznetsova et al. (2010). 
The mechanisms of pacemaker firing in LC neurons are not
fully understood, although there is the possibility that it is sustained by
a TTX-insensitive persistent sodium current (De Carvalho et al., 2000;
Alvarez et al., 2002). For serotonergic neurons of the DRN, there have
been no reports of a persistent sodium current and L-type calcium
currents are relatively small or absent (Penington et al., 1991) so the main candidate
for depolarization underlying pacemaking is a combination
of T-type calcium current and  the classical
fast TTX-sensitive sodium current which dominates the pre-spike
interval (Tuckwell and Penington, 2014).  In some cells the hyperpolarization-activated cation current may also play a role. 

In Figure 1 are shown portions of spike trains in NA neurons of mouse and rat LC, SE neurons in 
rat DRN and CRN and DA neurons in rat midbrain. It is noteworthy that the frequency of firing of LC and DRN 
principal neurons depends on sleep stage. Thus for example,  in rats, waking,
slow-wave sleep and REM sleep are accompanied by LC firing rates
of about 2.2 Hz, 0.7 Hz and 0.02 Hz, respectively (Foote et al., 1980;
Aston-Jones and Bloom, 1981; Luppi et al., 2012). 
  \begin{figure}[!h]
\begin{center}
\centerline\leavevmode\epsfig{file=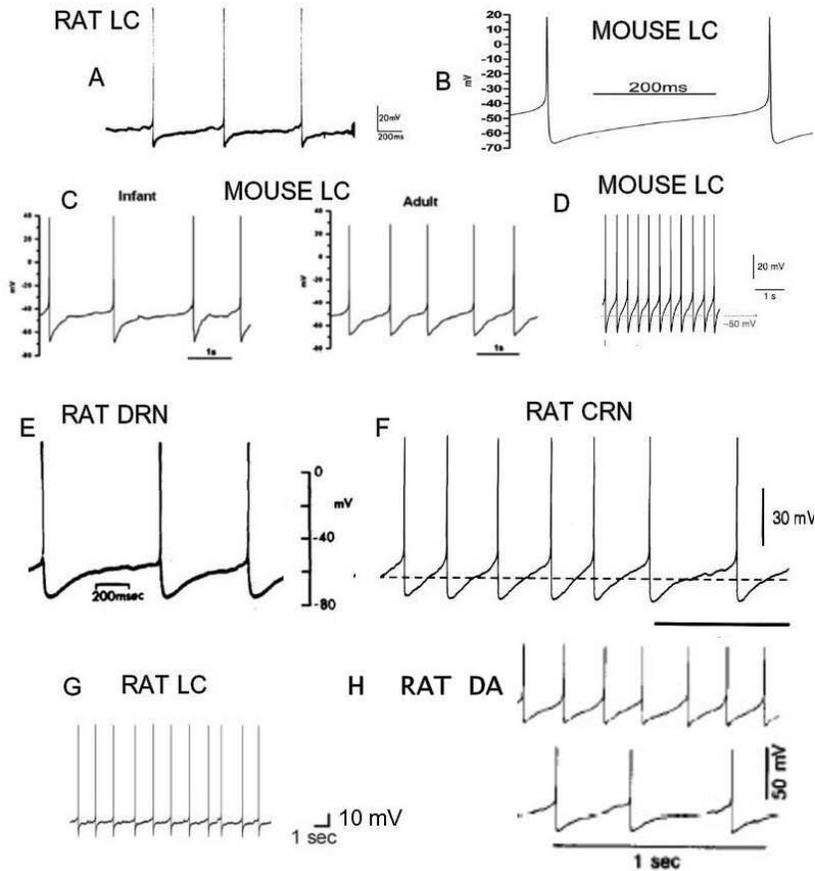,width=4.25in}
\end{center}
\caption{Some representative spikes from rat and mouse raphe nuclei and LC neurons.
{\bf A}. Part of a train of spikes in a rat LC neuron in slice. Markers 20 mV and 200 ms. (Andrade and Aghajanian, 1984). {\bf B}.  Detail of the course of the average membrane potential
in a mouse LC neuron during an interspike interval. (De Oliveira et al., 2010). {\bf C}. Action potentials in infant  (7 to 12 days) and adult (8 to 12 weeks) mice.
(De Oliveira et al., 2011). {\bf D}. Whole-cell current-clamp recording of spikes in a
 mouse (21 to 32 days) LC neuron. (Sanchez-Padilla et al., 2014). {\bf E}.
A few spikes from rat dorsal raphe nucleus (slice), (Vandermaelen and Aghajanian, 1983). {\bf F}. Portion of a spike train from rat caudal raphe nucleus. (Bayliss et al., 1997).  {\bf G}. Train of spontaneous spikes at a mean frequency of 0.85 Hz for a rat LC
neuron in vitro. (Jedema and Grace, 2004).  {\bf H}. In vitro spiking in
a rat dopaminergic neuron with two levels of applied current. (Grace and Onn, 1989).  } 
\label{fig:wedge}
\end{figure}

\newpage

Figure 2 shows  computed spikes for the detailed model of rat DRN SE neurons
 of Tuckwell and 
Penington (2014) with 4 different parameter sets. In this model the main
variables are membrane potential and intracellular calcium ion concentration
which satisfy ordinary differential equations. There are, however, 
11 membrane currents which drive the model, which results in a system with
18 components and over 120 parameters. In order to study quantities like 
interspike interval distributions with various sources of random synaptic input,
it is helpful to have a simpler system of differential equations which might
yield insight into the properties of the complex model whose execution
with random inputs over hundreds of trials would be overly time consuming.
It is also pointed out that LC and DRN principal neurons are each
responsive to activation of about 20 different receptor types,
making computational tasks even more cumbersome with an 18-component model
(Kubista and Boehm, 2006; Maejima et al., 2013). 
A simplified model of spike generation would be useful in modeling the
dynamics of serotonin release and uptake (Flower and Wong-Lin, 2014) and in simplified models of brainstem neural networks (Joshi et al., 2017). This approach 
has been proven useful by several authors, some of whom reduced the number of
component currents (for example Fitzhugh, 1961; Nagumo et al., 1962; Destexhe et al., 2001) whereas others have simplified the geometry of the dendritic tree (Rall, 1962  ; Walsh and Tuckwell, 1985; Bush and Sejnowski, 1993).

      \begin{figure}[!h]
\begin{center}
\centerline\leavevmode\epsfig{file=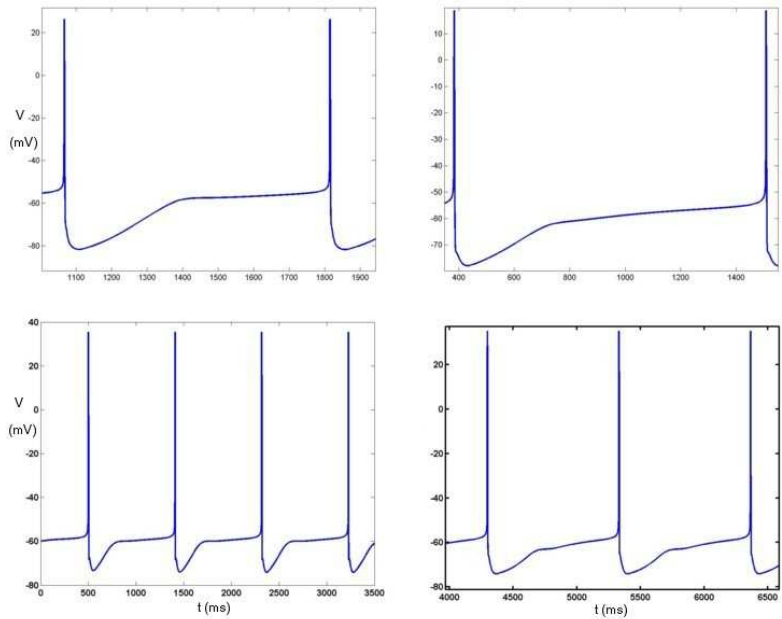,width=4.5in}
\end{center}
\caption{Examples of computed spikes in a model for serotonergic neurons of the rat
dorsal raphe nucleus from the model of Tuckwell and Penington (2014).
Illustrated are the prolonged afterhyperpolarizations after a spike. The
subsequent climb to threshold is plateau-like, sometimes being almost
horizontal. Membrane potentials in mV are plotted against time in ms.} 
\label{fig:wedge}
\end{figure}

\section{A reduced physiological model with two component currents}
With the multi-component neuronal model for DRN SE neurons (Tuckwell and Penington, 2014) there are 11 component currents and a variety of solution behaviors for different choices of
parameters. It is often difficult to
see which parameters are responsible for various spiking properties. 
Some parameter sets lead to satisfactory spike trains with 
properties similar to experiment, but sometimes spike durations are
unacceptably long.  Other sets give rise to 
spontaneous activity but with spikes which have uncharacteristic large notches on
the repolarization phase. Doublets and triplets are also often 
observed, and although these are sometimes found in experimental spike trains,
it is desirable to find solutions depicting the regular singlet spiking and
relatively smooth voltage trajectories usually observed.

Here we  describe a simplified two-component
model whose solutions exhibit membrane potential trajectories
 resembling those for pacemaker spiking in many brainstem neurons,
as depicted in Figure 1. The several currents which act to mainly depolarize
the neurons are lumped together in a single current denoted by $I_e$,
and similarly the currents which act to repolarize the cells are lumped into $I_i$.  Additionally there is an applied current $I_a$ which 
may be required to maintain pacemaker activity and which may also
contain excitatory and inhibitory synaptic input. $I_e$ and $I_i$
also subsume a leak current. 
The basic differential equation for the voltage $V(t)$ at time $t$ is written
\be  C\frac{dV}{dt}=-[I_e  + I_i + I_a] \ee 
where $C$ is capacitance.  The initial value of $V$ is usually taken to be
the resting potential, $V_R$. 
All potentials are in mV, all times are in ms, all conductances are in $\mu$S  and  capacitances are in nF.

The form of the current $I_e$ is based on that of the usual fast  sodium transient currrent
 $I_{Na}$  in the 11-component model for DRN SE cells and in the original
Hodgkin-Huxley (1952) model. The current $I_i$ is based on the form of  the
delayed rectifier potassium current $I_{KDR}$ in the 11-component
model. The inclusion of just the currents $I_e$, $I_i$, and
if necessary a small constant driving applied current, is found, with appropriate sets of parameters,  to be 
sufficient to give spikes whose trajectories resemble many found in pacemaker
spiking of brainstem neurons as exemplified by those illustrated
in Figures 1 and 2. These solutions have properties contrasting with
the usual ones for the Hodgkin-Huxley system - see for example
Hodgkin and Huxley (1952),
Tuckwell and Ditlevsen (2016) and Figure 10 herein.

The depolarizing current,  $I_e$ is given by the classical form 
\be I_e=g_{e,max} m_e^3h_e(V-V_e) \ee
with activation variable $m_e$, inactivation variable $h_e$, 
maximal conductance $g_{e,max}$ and reversal potential $V_e$. 
For the steady state activation we put 
\be m_{e, \infty} = \frac{1}{1 + e^{-(V - V_{e_1})/k_{e_1} } }. \ee
If the corresponding time constant is assumed to be voltage-dependent, we write it as 
\be \tau_{m, e}=    a_e + b_e e^{- \big( (V - V_{e_2})/k_{e_2}\big)^2}. \ee
The steady state inactivation is given by
\be h_{e, \infty}= \frac{1}{1 + e^{(V -V_{e_3})/k_{e_3} } }, \ee
and if its time constant is voltage dependent, we put 
\be \tau_{h, e}=   c_e + d_e e^{- \big( (V - V_{e_4})/k_{e_4}\big)^2}. \ee

We take the hyperpolarizing 
 current to be   
\be I_i= g_{i,max} n^{n_k}(V-V_i) \ee 
where $n$ (the traditional symbol)  is the activation variable,
$n_k$ is a usually positive integer-valued index,
$g_{i,max}$ is the maximal conductance and $V_i$ is the reversal potential.
The steady state activation for $I_i$  is written 
 \be n_{\infty}=\frac{1} { 1 + e^{- (V - V_{i_1})/k_{i_1}} }  \ee
and the time constant is, if voltage dependent, 
 \be \tau_n= a_{i} +  \frac {b_{i}} { \cosh( (V - V_{i_2})/k_{i_2}   )}   \ee
Activation and inactivation variables for $I_e$ are determined by the 
first order equations
\be \frac{dm_e}{dt} = \frac{m_{e,\infty} - m_e}{\tau_{m_e}}, \ee
\be \frac{dh_e}{dt}=\frac{h_{e,\infty} - h_e}{\tau_{h,e}}. \ee
The activation variable $n$ for $I_i$ satisfies an equation like (10).

In some instances, the time constants are chosen to be
constant, independent of V,  in which case they are denoted by $\tau_{m,e_c}$, $\tau_{h,e_c}$ and $\tau_{n,c}$ for $\tau_{m,e}$, $\tau_{h,e}$ and $\tau_n$, respectively.

\subsection{Examples of pacemaker-like firing}

We illustrate the pacemaker-like solutions with two choices of parameter sets.  Set 1 is based on the parameters used for fast sodium and delayed rectifier potassium in the 11-current component model  (Tuckwell and Penington, 2014). Set 2  is adopted from the parameters in a model of a rat sympathetic neuron with five conductances (Belluzzi and Sacchi ,1991). 
For the second set we also use the resting potential 
and the cell capacitance from Kirby et al. (2003) for rat DRN SE cells. The two sets of parameters
are summarized in Table 1.

\begin{center}
\begin{table}[!ht]
    \caption{Two basic parameter sets}
\smallskip
\begin{center}
\begin{tabular}{lcc}
  \hline
 Parameter &  Set 1 &  Set 2 \\
\hline 
    $V_{e_1}$ &   -33.1 &   -36\\ 
 $k_{e_1}$ & 8 & 7.2\\
 $V_{e_3}$ & -50.3 & -53.2 \\
 $k_{e_3}$ & 6.5 & 6.5 \\
$V_R$ & -60 & -67.8 \\
$\tau_{m,e_c}$  & 0.2 & 0.1 \\
$\tau_{h,e_c}$   & 1.0 & 2.0 \\
$C$ & 0.04 & 0.08861\\
 $V_{i_1}$& -15 & -6.1\\
 $k_{i_1}$& 7.0 & 8.0\\
$n_k$ & 1 & 1 \\
   $\tau_{n,c}$    & - & 3.5 \\
 $a_i$  & 1 & - \\
 $b_i$ & 4 & - \\
$V_{i_2}$ & -20 & -\\
 $k_{i_2}$ & 7 & - \\
$g_{e,max}$& 2.00& 1.5\\
$g_{i,max}$ & 0.5 & 0.5\\ 
$V_e$ & 45  & 45 \\
$V_i$ & -93 & -93 \\
  \hline
\end{tabular}
\end{center}
\end{table}
\end{center}

    \begin{figure}[!h]
\begin{center}
\centerline\leavevmode\epsfig{file=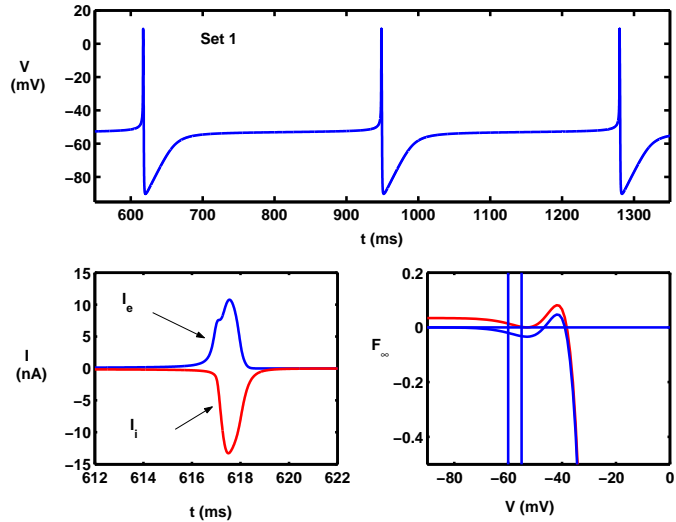,width=3.5 in}
\end{center}
\caption{Top. Repetitive spiking in the two-current ($I_e-I_i)$) model for the
parameter set 1 at the approximate threshold for spiking. 
Bottom left. The currents $I_e$ and $I_i$ during spikes.
Bottom right.  The function $F_{\infty}(V)$ defined in Equ. (12) without added current (thin blue curve), whereby spiking does not occur,  and with sufficient depolarizing current of 0.0342 nA (red curve) to give rise to pacemaker
activity. } 
\label{fig:wedge}
\end{figure}

Both of these  parameter sets led to repetitive spiking with the addition
of a small depolarizing current (see Table 2).  Typical spike trains are shown in Figures 3 and 4, 
and Table 2 contains lists of some of the details of the 
spike and spike train properties. Spiking for the second parameter 
set has a lower threshold for 
(repetitive) spiking, a longer ISI at threshold, a longer spike duration and a  larger
spike amplitude. For both parameter sets, most of these spike properties are in the ranges observed for
DRN SE neurons.
Figures 3 and 4 show well defined spikes with abruptly falling repolarization
phases to a pronounced level of hyperpolarization followed by a 
steady increase in depolarization until an apparent spike threshold is reached.
    \begin{figure}[!h]
\begin{center}
\centerline\leavevmode\epsfig{file=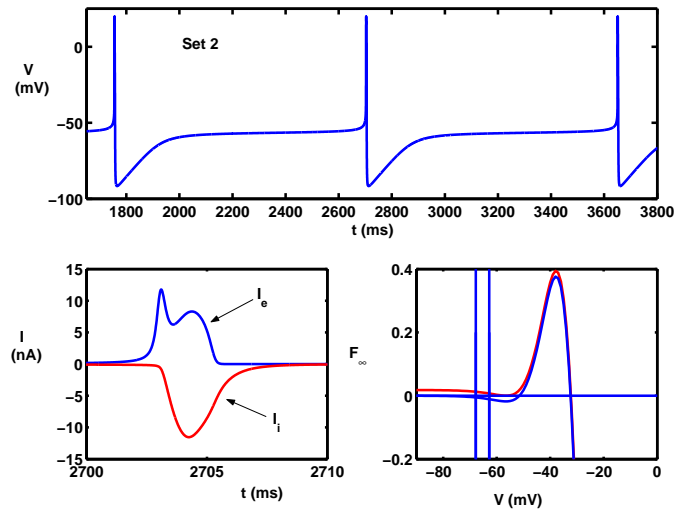,width=3.5 in}
\end{center}
\caption{Top. Repetitive spiking in the two-component model for the
parameter set 2 at the approximate threshold for spiking. 
Bottom left. The currents $I_e$ and $I_i$ during spikes.
Bottom right.  The function $F_{\infty}(V)$ defined in Equ. (12) without added current (thin blue curve), whereby spiking does not occur, and with sufficient depolarizing current of 0.018 nA (red curve) to give rise to pacemaker
activity} 
\label{fig:wedge}
\end{figure}
In the lower left-hand panels of Figures 3 and 4 are shown, on an expanded time scale,  the excitatory current $I_e$ and the inhibitory  current
$I_i$ during spikes. In the lower right-hand panels are shown
plots (thin blue curves) of the function $F_{\infty}(V)$ defined as the sum of the steady state ($t \rightarrow \infty$) values of the quantities of Equ. (2) and Equ. (7),
\be F_{\infty}(V) = -[g_{e,max} m_{e,\infty}^3(V)h_{e, \infty}(V)(V-V_e)
    +   g_{i,max} n_{\infty}^{n_k}(V)(V-V_i)]. \ee
The behavior of this function near the resting potential has been
found to provide a heuristic indicator for the occurrence of spiking (Tuckwell, 2013;
Tuckwell and Penington, 2014).  It can be seen in both Figures 3 and 4 
 that, for both parameter sets,  the function  $ F_{\infty}(V)$ is negative
for $V$ in an interval of considerable size around the resting potential
$V_R$, which is indicated by the vertical at -60 mV in Figure 3 and at
- 67.8 mV in Figure 4.  This means that around $V=V_R$
the derivative of $V$ with respect to time tends to be negative so that spontaneous
spiking is unlikely. The magnitude of the smallest depolarizing current required
to enable spiking is approximately the amount $-\mu_c$ which must
be added to make   $F_{\infty}(V) - \mu$ positive at and around
$V_R$. The resulting curves, shown in red in Figures 3 and 4,
are approximately tangential to the V-axis, being  obtained with 
$\mu_c=-0.0342$ for set 1 and $\mu_c=-0.018$ for set 2. Hence these
values of $\mu$ are estimates of the threshold depolarizing current
required for (pacemaker) spiking. 

Many brainstem (particularly DRN SE neurons) often have characteristically long plateau-like
phases in the latter part of the ISI and this is apparent in Figures 3 and  4 
for spikes elicited near the threshold for spiking for both sets of parameters.
The plateau for the second set is nearly three times as long as that 
for the first set. 
For both parameter sets,  (not shown) at a particular
value of $\mu$ the frequency jumps from zero to a positive value,
 being 3.0 Hz for set 1
and 1.1 Hz for set 2, so that 
this model with the chosen parameters would  be classified as
one with Hodgkin (1948) type 2 neuron properties. 

Figure 5 shows, for both parameter sets, 
the computed spike trajectories and ISIs for levels
of excitation not much above threshold, $\mu$ being from 
1 to 1.05 times the threshold values of $\mu_c$. 
In each case the spike trajectory displays a typical pronounced 
and prolonged post-spike afterhyperpolarization followed by
a plateau-like phase before the next spike.

The spiking properties for the two-component model  with the parameter
sets in Table 1 have
several features in common with the experimental 
spike trains of many brainstem neurons including DRN SE and LC NA cells.  The frequency of action potentials is  in good agreement 
with experimental values near threshold. However,  as the level of excitation
is increased to much greater values,  the frequency becomes 
somewhat high (not shown) relative to the most commonly reported  values for sustained firing in these brainstem neurons, although there are exceptions; for example, 
in rat LC neurons, Korf et al. (1974) found in unanesthetized in vivo preparations,  frequencies up to 30 Hz, 
and Sugiyama et al. (2012) reported in vivo frequencies over 7 Hz.
In midbrain serotonergic raphe neurons, Kocsis et al. (2006) obtained in vivo rates with mean 5.4 Hz. For experiments with depolarizing
current injection, Li and Bayliss (1998) obtained initial firing rates
of around 8 Hz for caudal raphe with an injected current of 60 pA;
Li et al. (2001) reported firing of rat DRN SE cells at frequencies as
high as 35 Hz with current injection, and Ohliger-Frerking et al. (2003)
found rates as high as 8 Hz and 11 Hz with 100 pA current injection in lean and obese
Zucker rats, respectively. See also subsection 4.1 on firing rates.

     \begin{figure}[!t]
\begin{center}
\centerline\leavevmode\epsfig{file=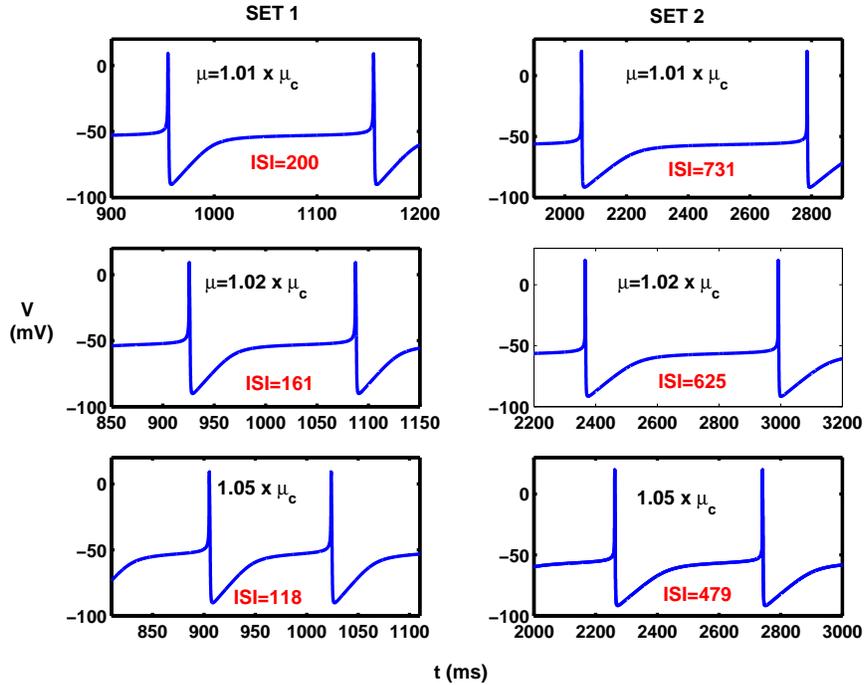,width=4.5 in}
\end{center}
\caption{Spike trajectories and ISIs obtained  in the reduced model for the
parameter sets of Table 1. Shown are results for Set 1 (left part) and Set 2 (right part), when
applied currents are increased by relatively small amounts of 1\%. 2\% and 5\% above threshold (top, middle, bottom curves).} 
\label{fig:wedge}
\end{figure}

\begin{center}
\begin{table}[!h]
    \caption{Spike train properties}
\smallskip
\begin{center}
\begin{tabular}{lcc}
  \hline
 Property &  Set 1 &  Basic set 2 \\
\hline 
 Spike threshold $\mu_c$   &  -0.0342   &   -0.018\\
ISI at threshold & 331 ms & 948 ms \\
Spike duration (-40 mV) &  1.6 ms  & 2.9 ms\\
Max V            &   +8  & +19.4 \\
Min V  &   -90.0  & -91.2 \\
  \hline
\end{tabular}
\end{center}
\end{table}
\end{center}

The simplified model is expected to be useful 
in predicting approximate responses to random synaptic inputs in contradistinction
to a sustained depolarizing current as employed in some 
experiments,  which can lead to very large firing rates.  
However, with other parameter sets the model may not  
exhibit such high frequencies as depolarization level increases.

As Figure 1 shows, not all regular spiking in brainstem neurons
entails a long afterhyperpolarization and  a long plateau. Figure 6
shows that spikes similar to those in Figures 1B, 1D and 1F are generated in the second model with increased depolarizing current.


    \begin{figure}[!ht]
\begin{center}
\centerline\leavevmode\epsfig{file=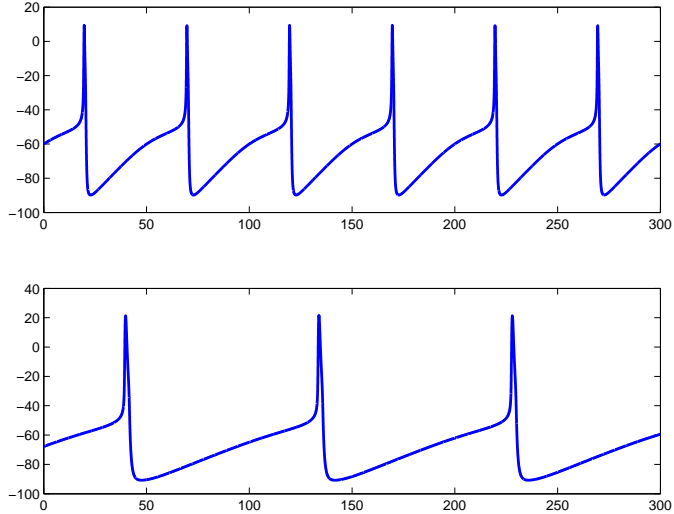,width=3.5in}
\end{center}
\caption{Spiking in the reduced model for the
parameter sets of Table 3 where in  both cases
$\mu=-0.05$, a value considerably above threshold and which 
results in short ISIs, characterized by relatively short plateaux and 
short-lasting AHPs.  Top part, set 1; bottom part, set 2.} 
\label{fig:wedge}
\end{figure}

\section{Effects of varying parameters}
The simplified model has 19 parameters for set 1, which compares with over 120 for the original model.  The dependence of spiking on
changes in the 19 parameters was investigated by computing solutions
with each parameter above and below its standard value as given in Table 1.  The focus in this subsection is the effect of parameter changes on the ISI, athough 6 other properties were routinely computed on each run, being spike duration, upslope on leading edge of spike, duration of AHP, amplitude of AHP, duration of plateau phase and approximate voltage at the half-way point of the plateau. 
The MATLAB computer programs  employed to determine these properties
are available from the authors on request. 

The \% change in the ISI for increases and decreases in  parameters
by $\pm$ 0.1 \% is shown for 19 parameters in Table 3. The entries are ranked according to the degree to which they change the ISI, with rank 1 for the parameter with the greatest effect. Examination of these results reveals that small changes in 6 parameters, ranks 1 to 6, can result in a very large change in the ISI, making this quantity infinite as
spiking is abolished altogether. Five of these 6 parameters are involved in the activation and inactivation of $I_e$ and the activation of $I_i$, the other one being the reversal potential for $I_i$. The inhibitory conductance, $g_{i,max}$ is the 7th ranked parameter making changes of 47.0 \% and -17.8 \% for increases and decreases 
of 0.1 \%, respectively.   Small changes in the parameters, ranked 8 to 11,  make moderate changes in the ISI of order 10 to 20 \%, being
the excitatory conductance, the depolarizing drive current $\mu$ and the excitatory reversal potential.  Changing the remaining parameters 
by $\pm$ 0.1\% has minor to insignificant effects on the ISI. The parameter whose such relatively small changes have the least effect on the ISI is the resting potential, $V_R$.

\begin{center}
\begin{table}[!h]
    \caption{Percentage changes in ISI for $\pm 0.1 \%$ changes  in parameter}
\smallskip
\begin{center}
\begin{tabular}{llcc}
  \hline
 Rank & Parameter  & +0.1 \% &  -0.1 \%\\
\hline 
 1 &  $V_{e1}$   &  $\infty$   &   -48 \\
3 & $k_{e1}$ & -40.2 & $\infty$ \\
6 & $V_{e3}$  & -26.6 & $\infty$ \\
11 & $k_{e3}$ & 2.7  & -2.4 \\
5 & $V_{i1} $ & -28.7 & $\infty$ \\
2 & $k_{i1}$ & $\infty$ &  -44.4 \\
19 & $V_R$ & 0.0 & 0.0 \\
4 & $V_i$ & -30.3 &  $\infty$ \\
 15 &  $\tau_{m,e_c}$  & 0.009  & -0.009 \\
16 & $\tau_{h,e_c}$   & -0.007 & 0.010 \\
13 &  $a_i$ & -0.03 & 0.03 \\
18 & $b_i$ & -0.002 & 0.002 \\
17 &  $V_{i2}$ &  0.007 & -0.007 \\
14 & $k_{i2} $ & -0.01 & 0.01 \\
8 & $g_{e,max}$ & -12.2 & 21.1 \\
7 & $g_{i,max}$ & 47.0 & -17.8 \\
9 & $\mu$ & 12.1 & -8.6 \\
12& $C$ & 0.14 & -0.14 \\
10 & $V_e$ & -6.3 & 8.0 \\
  \hline  
\end{tabular}
\end{center}
\end{table}
\end{center}

More details of the effects of changing the parameters are given
in Table 4. Here the parameters are grouped to reflect their
different strengths in changing the ISI. The results point to the
extreme sensitivity of the ISI to small relative changes in the following
8 parameters.
\begin{itemize}
\item{$g_{i,max}$, the maximal hyperpolarizing conductance. Increases of as
little as 0.005\% are sufficient to prevent spiking. Decreases of 0.05\% lead to a 43\% decrease in the ISI, or about a 50\% increase in firing rate.}
\item{$ V_{e1}$, the half-activation potential for the depolarizing current, $I_e$. When this parameter is increased by only 0.05\%, 
spiking is extinguished and when it is decreased by 0.05\% the ISI is
shorter by 37\%.}
\item{$k_{e1}$, the slope factor for the activation function of $I_e$. 
Decreasing this quantity by 0.05\% was sufficient to extinguish spiking,
whereas increasing it by the same amount led to a 28\% decrease
in the ISI.}
\item{$k_{i1}$, the slope factor for the activation of $I_i$.  Decreasing
this parameter by 0.05\% also extinguished spiking and an increase by 0.05\% gave a 33\% reduction in ISI, or a 50\% increase in firing frequency.}
\item{$g_{e,max}$, the maximal depolarizing conductance. A reduction of as little as 0.05\% ln this quantity was sufficient to abolish spiking, whereas a 0.05\% increase resulted in a 72\% decrease in the ISI.}
\item{$V_i$, the reversal potential for the hyperpolarizing current,
A decrease of 0.05\% of this parameter gave a 65\% increase in ISI,
equivalent to a substantial reduction in firing rate, whereas a 0.05\% increase led to a 20\% decrease in the ISI.}
\item{$C$, the whole cell capacitance. Reducing this quantity by
the relatively small amount of 0.05\% resulted in about a 50\% increase
in firing rate whereas a 0.5\% increase in $C$ led to a 50\% decrease in firing rate.}
\item{$\mu$, the added current. Increasing this by 0.05\% (that is
making it less negative and hence less depolarizing) made the ISI infinite, whereas decreasing $\mu$ by 0.05\% gave a 27\% reduction
in the ISI.}
\end{itemize}

\begin{center}
\begin{table}[!h]
 \caption{\% changes in ISI for various \% changes in set 1 parameters}
\smallskip
\begin{center}
\begin{tabular}{lcccccc}
  \hline

\% Changes $\rightarrow$   & 0.05 & 0.01 & 0.005 & -0.005 & -0.01 & -0.05 \\  
\hline 
$V_{e1}$     & $\infty$  & 25.7 & 10.4 & -7.6 & -13.6  & -37.0 \\  
$V_{e3}$  &  -16.9 & -4.2 & -2.3 & 2.4 & 5.1 & 41.5 \\
$V_{i1}$ &  -18.6& -4.8 & -2.5 & 2.8 & 6.4 & 53.7 \\
$k_{e1}$ & -28.6 & 9.1 & -4.9& 5.8 & 13.0 & $\infty$ \\
$k_{e3}$  & 1.3 & 0.3 & 0.1  & -0.1 & -0.3 & -1.2 \\
$k_{i1} $ & $\infty$ & 18.1 & 7.8& -6.1& -11.1 & -32.8\\
$V_R$ & 0.0 & 0.0 &0.0& 0.0 & 0.0 & 0.0 \\
$V_i$ & -19.8 & -5.3 & -2.8 & 3.1 & 6.5 & 65.1 \\
\hline
  \% Changes & 50 & 25 & 10 & -10 & -25 & -50 \\  
  \hline 
$\tau_{m,e_c}$  & 4.7 &  2.5 & 1.0 & -1.0 & -2.6 & -5.3\\
$\tau_{h,e_c}$ & -4.2 & -2.1 & -0.8 & 0.7 & 1.8 & 1.8 \\
\hline
\% Changes & 50  & 20 & 10 & -10 & -20 & -50 \\
\hline
$a_i$ & -17.2 & -9.3 & -3.4 & 3.3 & 6.6 & 15.4 \\
$b_i$ & -1.3 & -0.5& -0.2 & 0.3 & 0.5 & 1.3 \\
$V_{i2}$ & 1.6 & 1.0 & 0.6 & -0.8 & 1.9 & -7.8\\
$k_{i2}$ & -9.7 & -3.1 & -1.4 & 1.1 & 1.8 & 2.5\\
\hline
\% Changes & 10 & 5 & 1 & -1 & -5 & -10\\
\hline
$g_{e,max}$ & -72.4 & -66.7 & -46.1 & 8.9 & 21.1 & $\infty$ \\
$g_{i,max}$ & $\infty$ & $\infty$ & $\infty$ & -10.4 & -17.8 & -43.1\\
\hline
\% Changes & 0.5 & 0.1 & 0.05& -0.05 & -0.1 & -0.5 \\
\hline
$\mu$ & $\infty$ & 12.1 & 5.5 & -4.6 & -8.6 & -27.9 \\
\hline
\% Changes & 50 & 25 & 10 & -10 & -25 & -50\\
\hline 
$C$ & 66.2& 33.3 & 13.4 & -13.5 & -33.7 & -67.4 \\
\hline
\% Changes & 1 & 0.5 & 0.1 & -0.1 & -0.5 & -1 \\
\hline
$V_e$ & -33.1 & -22.3 & -6.3 & 8.0 & 111.8 & $\infty$ \\
\hline
\end{tabular}
\end{center}
\end{table}
\end{center}

    \begin{figure}[!h]
\begin{center}
\centerline\leavevmode\epsfig{file=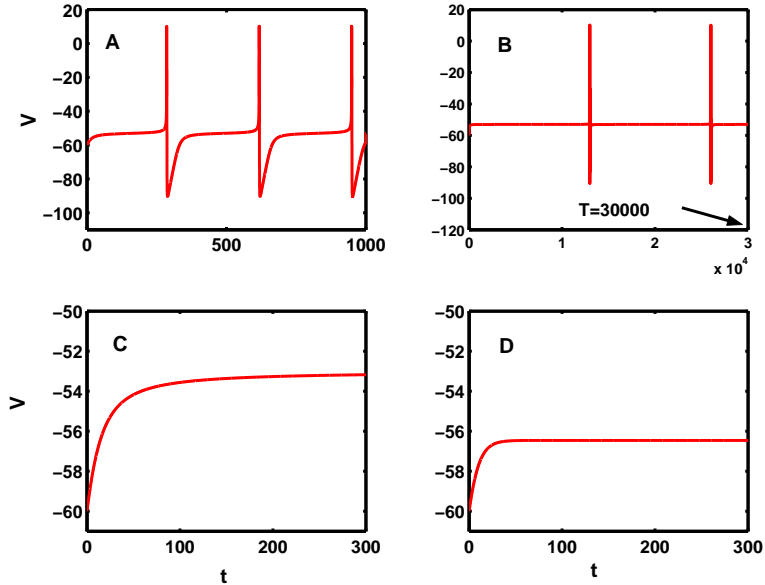,width=4in}
\end{center}
\caption{Voltage versus time plots for various values of the
half-activation potential of the excitatory current.  A. The standard set
of parameters including the value of $V_{e1}$ at -33.1 mV. Pacemaker activity is sustained with an ISI of 331 ms. B. The value of $V_{e1}$  is slightly above the critical value for spiking at 0.02480 \% above the standard value. The ISI is about 13000 ms.  C. With the value of
$V_{e1}$ slightly higher at 0.02481 \% above the standard value,
there is no spiking and $V$ approaches a limiting value of about 
-53 mV after about 200 ms. D. An even larger value of $V_{e1}$,  2.5\%
above -33.1 mV, results in an asymptotically smaller departure from rest to about -56 mV which is
achieved in about 20 ms. } 
\label{fig:wedge}
\end{figure}

    \begin{figure}[!h]
\begin{center}
\centerline\leavevmode\epsfig{file=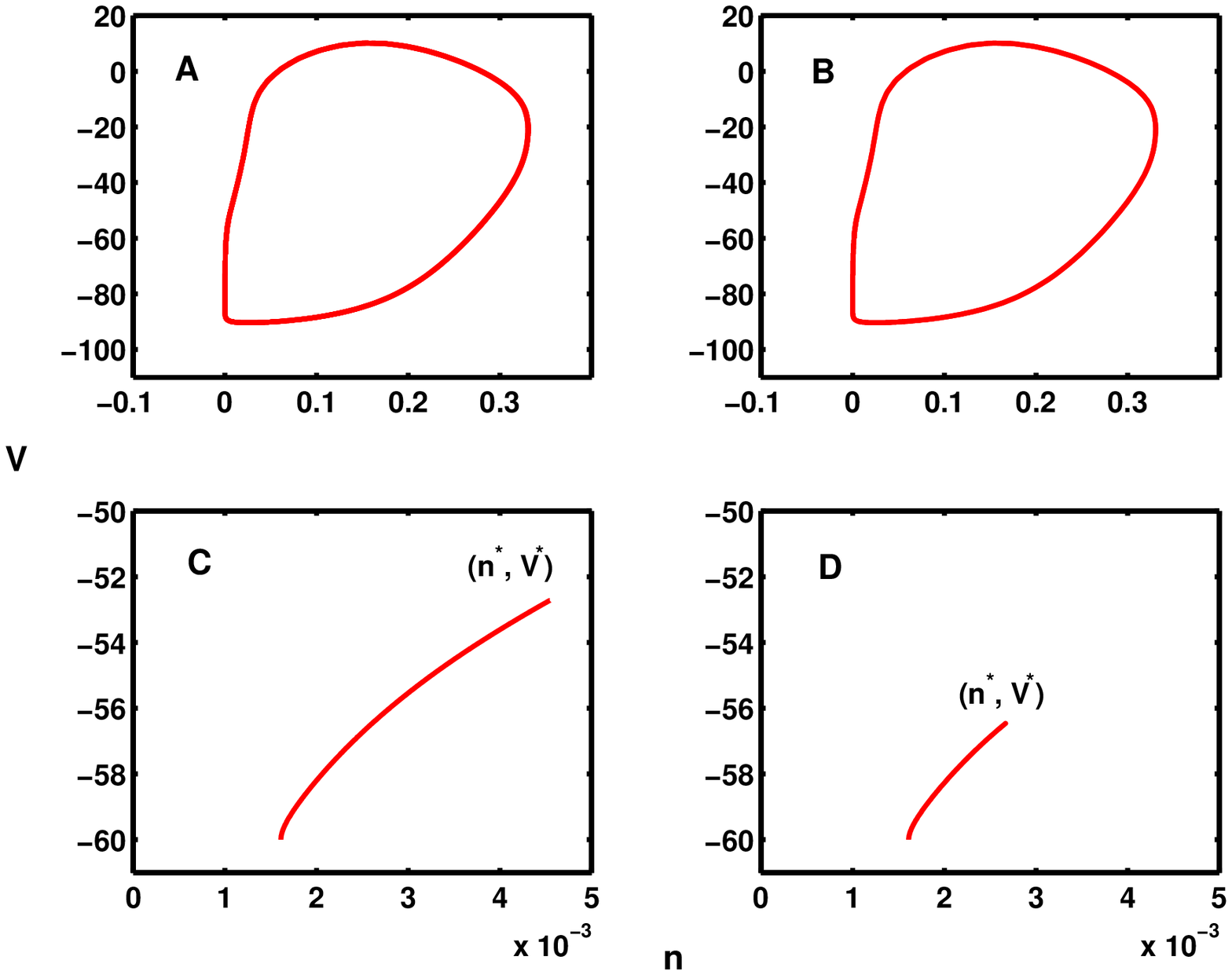,width=4in}
\end{center}
\caption{Plots of $V$ versus $n$ for the corresponding plots A-D of $V$ versus $t$ in Figure 7. Note the similarity of the trajectories in A and B, despite enormously different ISIs.  In C and D, corresponding to only subthreshold responses,  trajectories are shown approaching 
asymptotic values at $(V^*,n^*)$.  } 
\label{fig:wedge}
\end{figure}

Changes in the remaining 11 parameters have moderate to minor effects on the ISI - see also Table 3.

\subsection{More detailed analysis for $V_{e1}$ and $\mu$}
In this subsection the transition from non-spiking to spiking for
two of the parameters is examined more closely near the critical
values.
\subsubsection{Half-activation potential for $I_e$}
The parameter ranked 1 in Table 3 is $V_{e1}$ which is the half-activation potential for the activation variable of the excitatory current.
Its standard value was -33.1 mV which gave rise to pacemaker-type spiking with an ISI of 331 ms. Since only a small increase in this 
parameter abolished spiking, it is of interest to study the behavior of
solutions near a critical value above -33.1 at which no spiking occurs.
To this end, solutions were computed with set 1 parameters with 
varying values of  $V_{e1}$. Results are shown in Figures 7 and 8.
In Figure 7 are shown computed spike trains, i.e., $V$ versus $t$, whereas in Figure 8 are shown plots of $V$ versus $n$ for the indicated time periods.

In panel A, the value of $V_{e1}$ is the standard value (-33.1) and the spike train is as described above with a constant ISI (Figure 7A). The trajectories 
of $V$ versus $n$ for each spike are practically identical, motion being
clockwise on these orbits (Figure 8A). 

In the panels B, the value of $V_{e1}$ is just 0.024800  \% above the standard value and this is very close to the critical 
value (call it $V_{e1,c}$)  beyond which spiking does not occur because it is too far from
the resting potential. As seen in Figure 7B, the first spike occurs at 12980 ms and the second at 26010 ms, but somewhat remarkably, the orbits (for the two spikes)as seen in Figure 8B are visually indistinguishable from those in Figure 8A where the ISI 
is only about 1/40 as long at 331 ms. 

The $V$ versus $t$ curve  for a value of $V_{e1}$ 
which is slightly larger at  0.024810 \% above the standard value, and
presumably greater than the critical value  $V_{e1,c}$, is shown in
Figure 7C. Here there are no spikes (up to 30000 ms) and the values
of $(V,n)$ have limiting values $(V^*,n^*)$ of about (-53.0038, 0.004368) as seen in Figure 8C.  Again, somewhat remarkably, the trajectory from
rest to $(V^*, n^*)$ lies almost exactly on the spiking orbit as it
approaches  the  limiting value.

 With the considerably higher value of $V_{e1}$, 
at 2.5 \% above the standard value, there are, as expected, no spikes as seen in Figure 7D.  The equilibrium point is closer to rest at 
 $(V^*,n^*)$=(-56.4600, 0.002670).  The trajectory in $(n,V)$ space still adheres to the early part of the spike orbit.  

\subsubsection{Magnitude $\mu$ of the added current: comparison with usual Hodgkin-Huxley parameters}
The parameter ranked 9 in Table 3 is $\mu$ which is the magnitude
of the added depolarizing current which may
be required to induce pacemaker-type spiking. Its set 1 value
was $\mu=-0.0342$ which was near the threshold value for spiking
when all other parameters had their set 1 values. The solutions are not as sensitive to changes in this parameter as to those in  $V_{e1}$.
Figure 9 shows the plots of $V(t)$ versus $n(t)$ for four values of $\mu$. 

 In the left panel orbits are shown for the the standard value -0.0342 of $\mu$, which gives an ISI of 331 (red curves), and for a value 0.44 \% above (more positive and hence more hyperpolarizing) the standard value (green curves) which yields an  extremely small spike rate 
with the first spike not occurring until 13,700 ms. The orbits for these two values of $\mu$ are almost coincident, but one can on close
inspection distinguish parts of the red and green curves.  The critical value of $\mu$, called $\mu_c$, for spiking is apparently not far above
the second value. 

In the right panel of Figure 9 are shown orbits, effectively from $t=0$
to $t=\infty$,  for two values of $\mu$ above $\mu_c$, at which there is no evidence of spiking. The red curve is for a value of $\mu$ just
0.5 \% above the standard value. The asymptotic value of $(V,n)$ is
with $V^*=-53.2313$ and $n^*=0.004229$. Again the curve closely adheres to the early part of the spike trajectory.  This is also the case
for the much higher value of $\mu > \mu_c$ at 20 \% above the standard value. The curve for this case is blue and terminates 
a shorter distance from the rest point, with $V^* \approx -57.21$ and $n^*=0.0024$. 

\u
\noindent {\bf Usual HH parameters}
\u
The manner in which the periodic orbit emerges as $\mu$ passes
through a critical value in the above calculations is seemingly quite different from that in which
periodic spiking arises in the HH model with the usual
parameters as given, for example,  in Tuckwell and Ditlevsen (2016).
The latter case is illustrated in Figure 10 where there are shown 
in A, B, C, respectively the 
responses to input currents well below threshold for periodic
spiking ($\mu=3.96$), not far below ($\mu=5.61$) and above threshold
($\mu=6.6$). Here the subthreshold responses are oscillatory and
decay is into the rest point. This is in contradistinction to the behavior
of solutions with the parameters of the present model - see Figure 9.
The nature of the bifurcation to periodic solutions that occurs at the critical values for the original
Hodgkin-Huxley parameters was discussed in Tuckwell et al. (2009).  
Further analysis is required to see if the bifurcation to periodic
solutions in the present model has a different mathematical structure
from that in the original model.

    \begin{figure}[!h]
\begin{center}
\centerline\leavevmode\epsfig{file=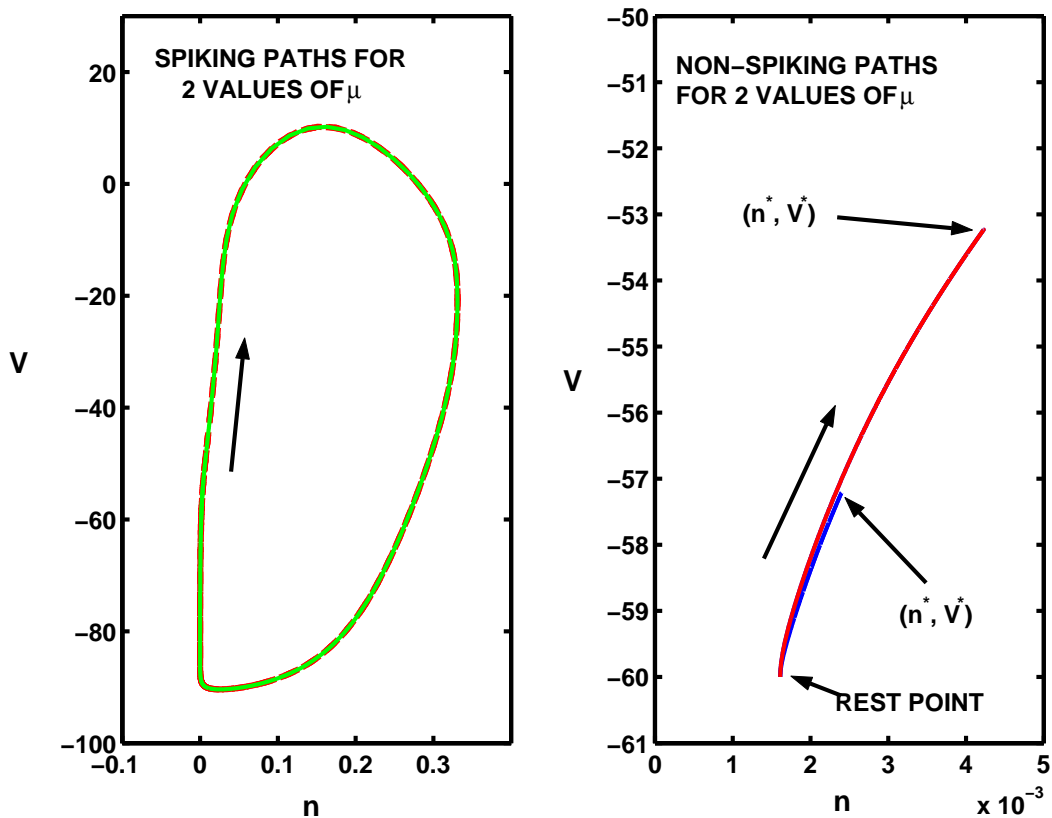,width=5.5in}
\end{center}
\caption{Plots of $V$ versus $n$ as the depolarizing current
is changed above and below the critical value for spiking.
In the left panel are shown spiking orbits for the standard value of $\mu$ =-0.0342 (red curves) which gives an ISI of 331 ms, and a value which
is 0.44 \% above the standard value (green curve) which results in an ISI of about 14,000 ms. The phase diagrams for these widely disparate 
values of the ISI are almost identical.  In the right hand panel are 
orbits for values of $\mu$ which are too large (hyperpolarizing) to
induce spiking. The orbits again asymptote to subthreshold values at
$(V^*,n^*)$. These subthreshold orbits coincide with the early parts of
the spiking orbits shown in the left panel.} 
\label{fig:wedge}
\end{figure}

    \begin{figure}[!h]
\begin{center}
\centerline\leavevmode\epsfig{file=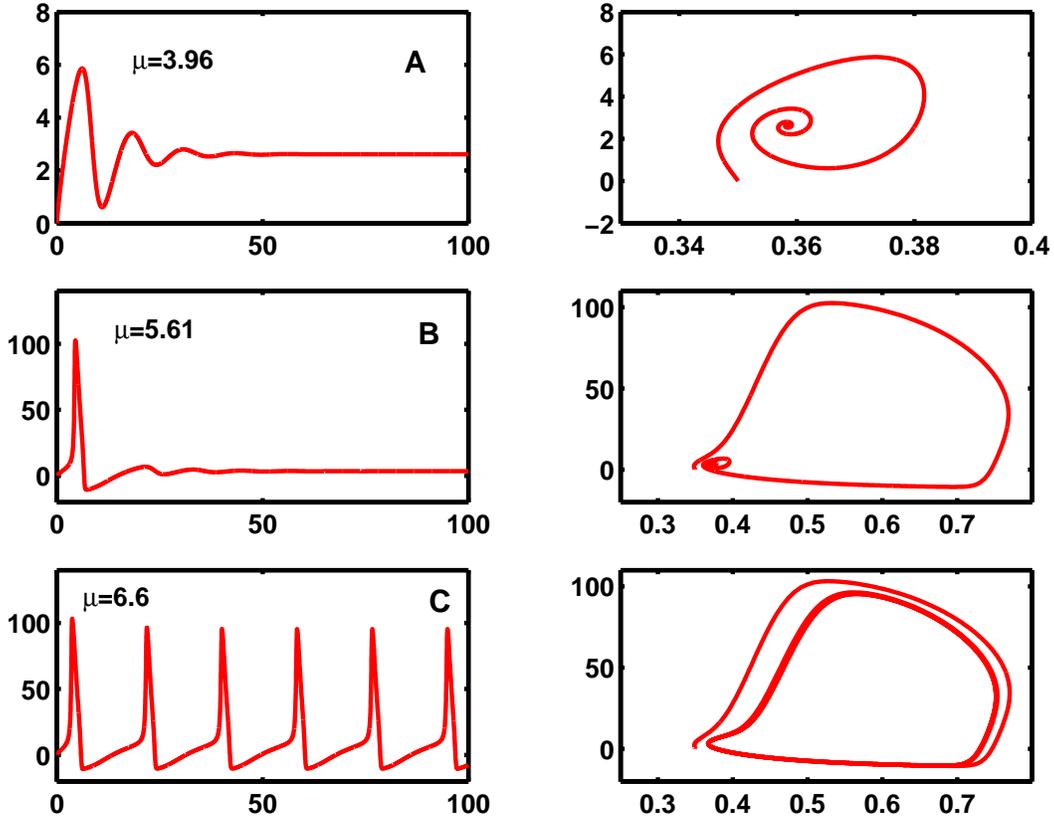,width=5.5in}
\end{center}
\caption{Solutions $V(t)$ versus $t$ (left column) and phase diagrams
$V$ versus $n$  (right column) for the Hodgkin-Huxley equations with the original parameters.  In A, the input current is well below the spiking threshold, in B it is not far below and in C, it is above the threshold
for repetitive spiking.  The subthreshold behaviors are quite different
from those illustrated in Figure 9, right hand diagram.} 
\label{fig:wedge}
\end{figure}

\subsection{Changing plateau level and the minimum of the AHP}
Examination of the membrane potential trajectories during pacemaker-type firing often shows a relatively long plateau between successive spikes during which the membrane potential increases slowly, sometimes very slowly before climbing past threshold to give rise to
the subsequent spike.  Such plateaux are exemplified by the recordings shown in Figure 11, where the dopamine neuron spikes are from Grace and Bunney (1995), those for
noradrenaline from  Andrade and Aghajanian (1984), and those for serotonin from Vandermaelen and Aghajanian (1983).  In the case of the in vitro recording shown for the DA cell, the plateau phase is 
not so flat, but an example with a flat plateau phase is shown in Figure 1H (from Grace and Onn, 1989)  for a cell with an applied hyperpolarizing current of 0.07 nA.  The resting potential for the DA cell shown in Figure 11 (middle dotted line) is -50 mV, for rat NA cells the average resting potential is -58.2 mV based on
11 sets of experimental results, and the average resting potential
for SE neurons of the DRN is about -60 mV. 
  
    \begin{figure}[!h]
\begin{center}
\centerline\leavevmode\epsfig{file=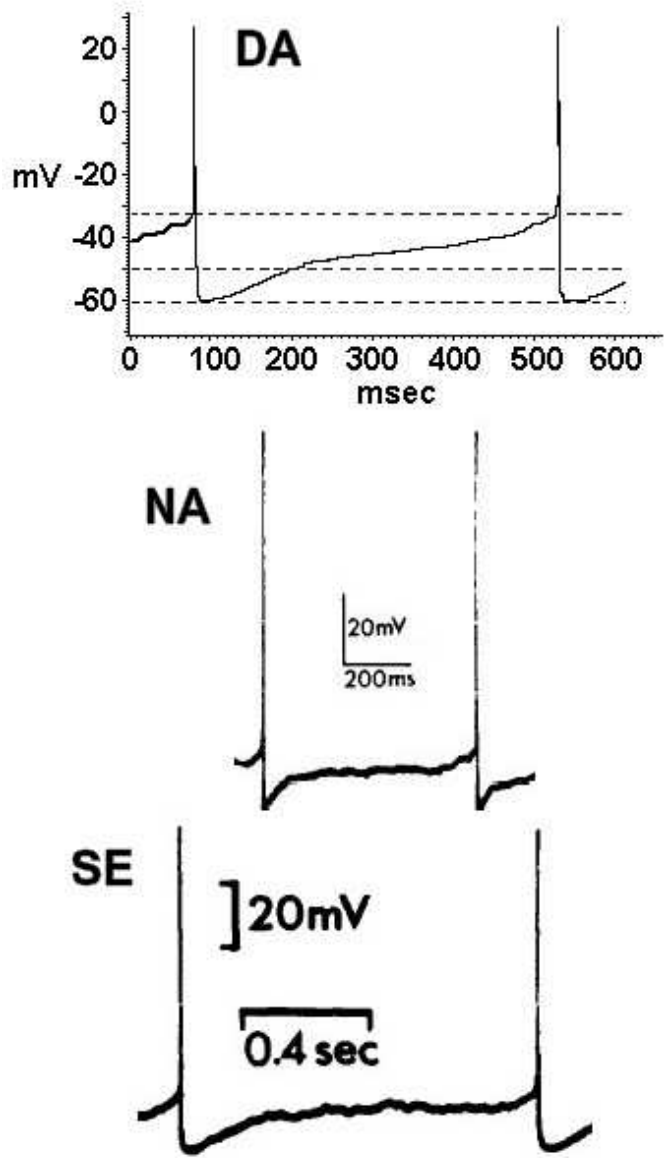,width=3.5in}
\end{center}
\caption{Pacemaker-type spiking in a DA neuron (top) from Grace and Bunney (1995) (also in Grace and Onn, 1989), from an NA neuron (middle) from Andrade and Aghajanian (1984), and an SE neuron (bottom) from Vandermaelen and Aghajanian, (1983). In each trajectory there is evidence of an
extended plateau phase during which the membrane potential undergoes relatively minor changes before ascending at the end of the ISI to give rise to a subsequent spike.} 
\label{fig:wedge}
\end{figure}

The DA and
NA recordings are in vitro recordings from rat substantia nigra and
locus coeruleus, respectively.  It is likely that regular pacemaker-type 
spiking in these two types of cells usually only occurs in vitro.
Grace and Onn (1989) stated that DA cells recorded in vitro fired
exclusively in a highly regular pacemaker-like pattern. This is 
supported by  results reported by  Grenhoff et al. (1988) for DA neurons and by Svensson et al. (1989) on NA neurons. In these two sets of in vivo experiments, application of the excitatory amino acid antagonist kynurenate (or kynurenic acid) changed the irregular (bursting) 
nature of spiking of DA and NA neurons to regular pacemaker
type activity. 

 Since in vitro recordings usually have diminished and
possibly no synaptic input, the findings in these latter two references
offer a plausible explanation why regular pacemaking often occurs in vitro but rarely in vivo
(see for example Sugiyama et al., 2011). Findings related to that of Grenhoff et al. (1988) were that the GABA$_B$ agonist baclofen
made firing in DA neurons slower and more regular (Engberg et al.,1993), and that similar effects resulted from the application of NMDA receptor blockers (Overton and Clark, 1992; Cherugi et al., 1993). In another more recent study, Blythe et al. (2009) found that in DA neurons of the substantia nigra, bursting could be induced by the application
of glutamate at somatic or dendritic sites. In a related interesting investigation in DA neurons, Putzier et al. (2009) found that the voltage 
characteristics of L-type \ac channels rather than \CA selectivity were important factors in pacemaking and bursting activity. 

 For SE neurons, regular pacemaking has been found in
both in vivo and in vitro experiments, suggesting less fine tuning
by excitatory synaptic inputs with a more robust pacemaker mechanism.
Levine and Jacobs (1992) applied kynurenic acid to cat DRN 
serotonergic neurons and found that their spontaneous firing rates were only reduced by about 4 \%.

It can be seen that the voltage levels of the plateaux between the NA and SE spikes in Figure 11 are often not much above resting potential. 
However for the model with set 1 parameters, as shown in Figures 3 and 5,   the plateau level is
about 6.8 mV above resting value. 
It turns out that the plateau level
and minimum can be easily adjusted with a change of parameters
which  are voltages, being in particular $V_e, V_i, V_{e1}, V_{e3}, V_{i1}, V_{i2}$. These were all changed together by an amount
$\Delta V$.  With $\Delta V=-5$ mV applied to set 1 parameters
no spikes emerged. 

With all set 1 parameters unchanged with the exception that the reversal potential for the inhibitory current was at a more
depolarized level of  $V_i=-88$ mV,  spiking was very rapid with an ISI of only 67.8 ms.  If the maximal inhibitory conductance was then multiplied by 1.1427, the ISI was back to 338.9  ms, close to the value for set 1 parameters.   Calling this set 3, with altered $V_i$ and altered
$g_{i,max}$, the plateau level was 7 mV above resting level. 
When the set  $V_e, V_i, V_{e1}, V_{e3}, V_{i1}, V_{i2}$ was 
uniformly decreased by 5mV, (set 3'), 
the ISI was virtually unchanged, but the plateau level was reduced to only 2mV above resting potential.  The spike trains for sets 3 and 3' are shown in Figure 12, with red and blue curves, respectively. 
The minimum value of $V$ during the AHP for set 3 is -85.7 mV
and -90.7 for set 3'. Some of the remaining spike train properties 
for set 3 parameters (with corresponding properties for set 3' in brackets) are: spike duration 1.6 ms (1.27), upslope of spike 100 V/s
 (111),
duration of AHP 38.8 ms (64), AHP amplitude 25.7 mV (30.7) and plateau duration 270 ms (247). Several of these properties are quite different for sets 3 and 3' and it is remarkable that the ISIs are almost  identical. However, it has been shown that the plateau level can be
easily adjusted to various values above resting level. 

    \begin{figure}[!h]
\begin{center}
\centerline\leavevmode\epsfig{file=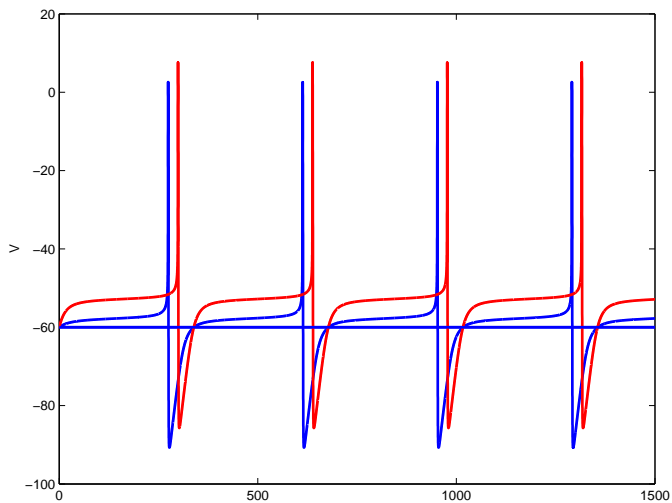,width=3.5in}
\end{center}
\caption{Adjustment of plateau level and voltage minimum during AHP in  pacemaker-type spiking in the 2-component model with 
parameters given in text (red curve) as well as with changes of $\Delta V=-5$ mV in the 6 key voltages 
$V_e, V_i, V_{e1}, V_{e3}, V_{i1}, V_{i2}$  (blue curve).} 
\label{fig:wedge}
\end{figure}

\section{Firing frequency and bursting}
\subsection{Firing rates}
Examination of the experimental literature on firing rates and 
their responses to applied currents for the three main types
of neuron, DA, NA, and SE,  considered above,
reveals a wide range of often disparate results. The origins of such
variability lie in several factors including animal types and ages and  
experimental conditions and techniques. Generally, in pacemaking mode, firing rates of locus coeruleus (NA) neurons (Tuckwell, 2017, contains a summary)  are less than 5 Hz with a few in vivo experiments reporting higher or
much higher rates. For serotonergic neurons of the DRN (summarized in Tuckwell, 2013), firing rates are usually less than a few Hz, rarely around 4-5 Hz and with applied current up to 10 Hz (Ohliger-Frerking et al., 2003).  Rates as high as 20 Hz in cat DRN were reported under glutamate application by Levine and Jacobs (1992).  Bursting is also sometimes reported in these cells (H\'aj\'os et al., 1996; H\'aj\'os et al., 2007). As remarked above, DA cells in vivo generally exhibit burst firing and in vitro pacemaking 
was found to have an average value of 5.4 Hz in one experiment
(Grenhoff et al., 1988).  Firing rates for set 1 parameters in the
simplified model were computed for both more hyperpolarizing
applied currents (Figure 13A) and more depolarizing currents (Figure 13B) relative to the standard value of 34.2 pA.  Frequencies ranged from 0 to
3 Hz in Figure 9A and from 3 to about 8 Hz in Figure 9B, the rather high frequencies in the latter resulting from large applied depolarizing currents. The frequencies for set 2 parameters were considerably less (see Figures
4 and 5).

  \begin{figure}[!h]
\begin{center}
\centerline\leavevmode\epsfig{file=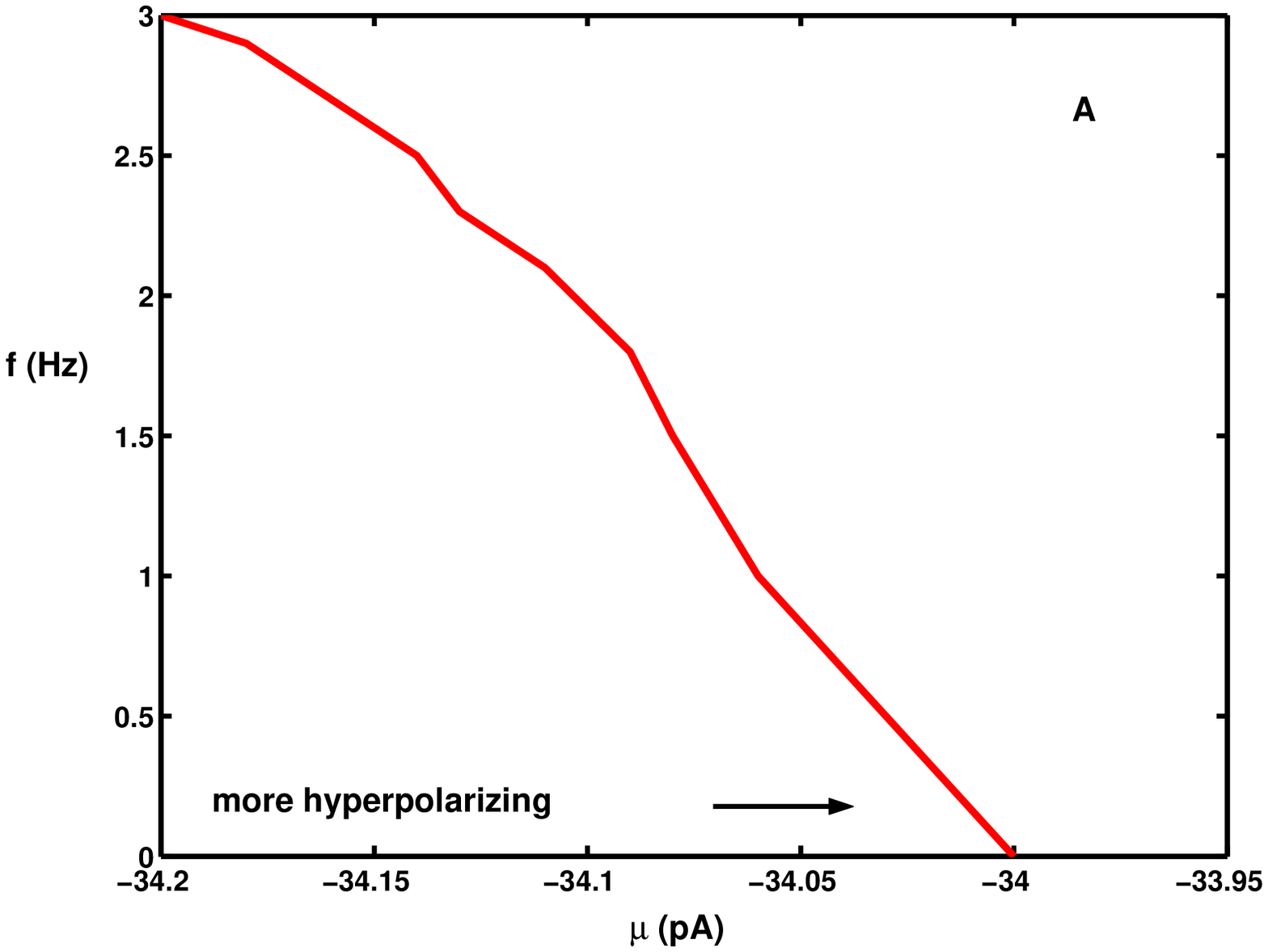,width=2.7in}\epsfig{file=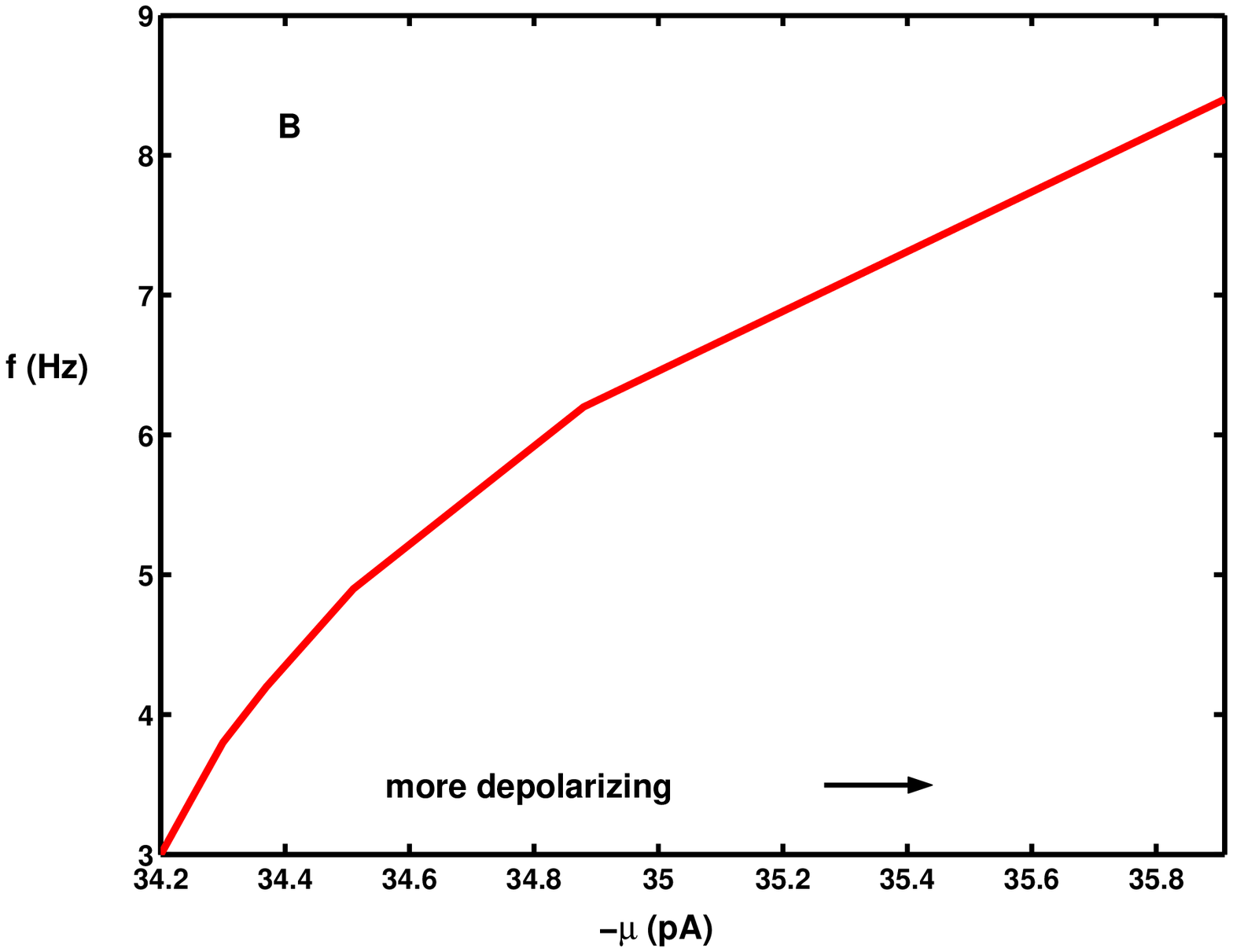,width=2.7in}
\end{center}
\caption{{\bf A}. Frequency of firing for set 1 parameters with values of
$\mu$ which are more hyperpolarizing than the standard value of $-0.0342$ nA.
{\bf B}. As in part A but that applied currents are more depolarizing than
$-0.0342$ nA. } 
\label{fig:wedge}
\end{figure}

\subsection{Bursting}
The results on the mechanisms of bursting in DA and LC neurons mentioned above  (Grenhoff et al.,1988;  Svensson et al., 1989; 
Overton and Clark, 1992;  Cherugi et al., 1993; Engberg et al.,1993;  Blythe et al., 2009) indicate that in several instances regular pacemaking
in these cells may be replaced by bursting if a (synaptic) excitatory drive is activated or an inhibitory drive inactivated. This was explored in the
simplified model by briefly perturbing the constant excitatory conditions that led to regular pacemaking. The perturbations mimic the effects of either increased excitatory input or decreased inhibitory input.  The results are shown in Figure 14.  In all cases the perturbations lead to
rapid spiking which mimics bursting over the duration of the perturbation with a subsequent return to pacemaker spiking when  the perturbation is terminated. In Figures 14A and 14B, the perturbations consist of
increases in the maximal excitatory conductance  $g_{e,max}$ for
200 ms (14A) or decreased inhibitory maximal conductance (14B). 
In Figures 14C and 14D, the depolarizing current parameter $\mu$
is increased for either 200 ms (14C) or 100 ms (14D). In all four cases
the evidence is convincing that such perturbations may lead to a disturbance of pacemaker firing. Thus, a steady but intermittent
flow of excitatory synaptic input or a reduction of inhibitory synaptic input
could lead to non-pacemaker firing and occasional or frequent bursting.

    \begin{figure}[!h]
\begin{center}
\centerline\leavevmode\epsfig{file=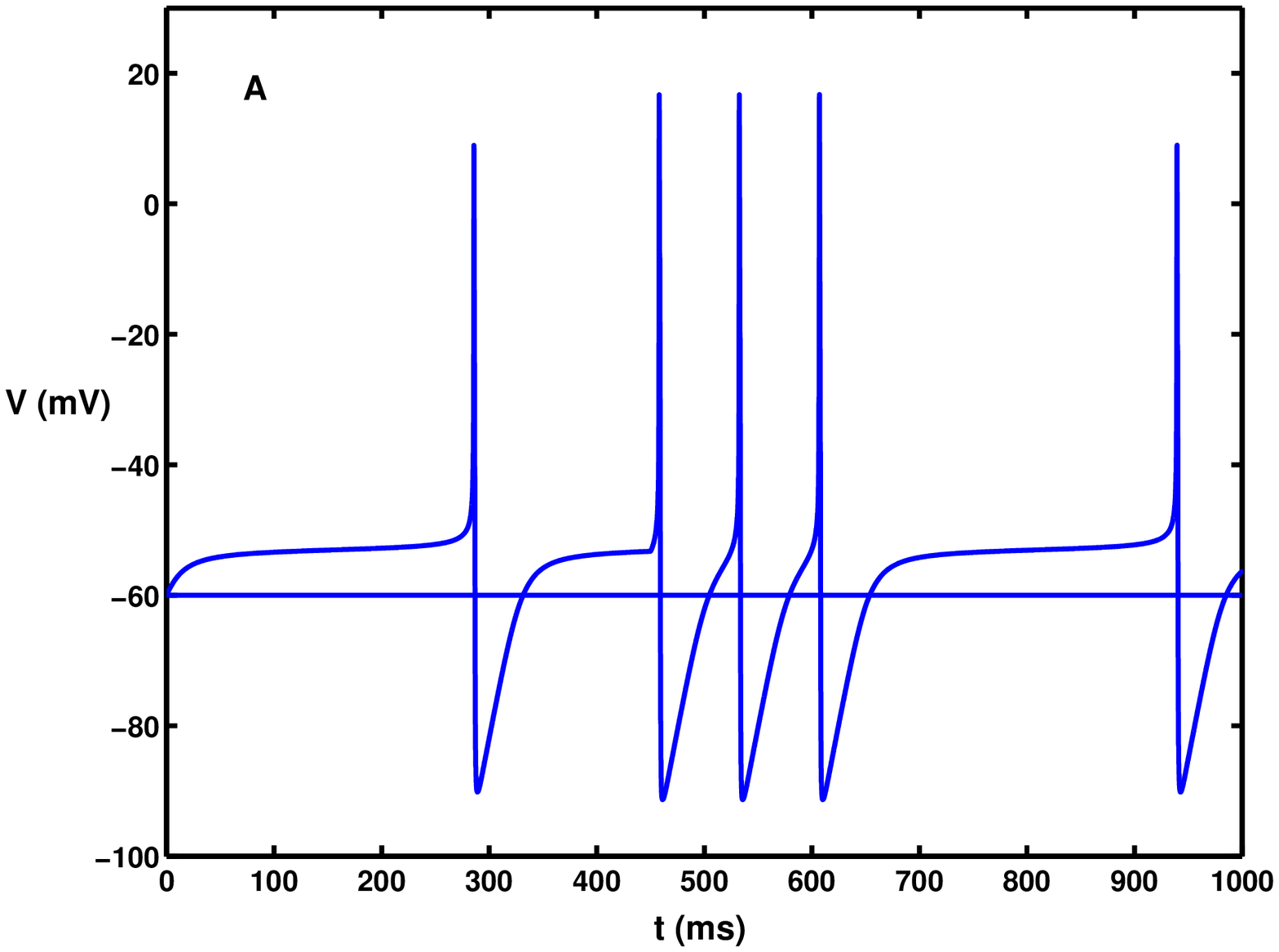,width=2.7in}\epsfig{file=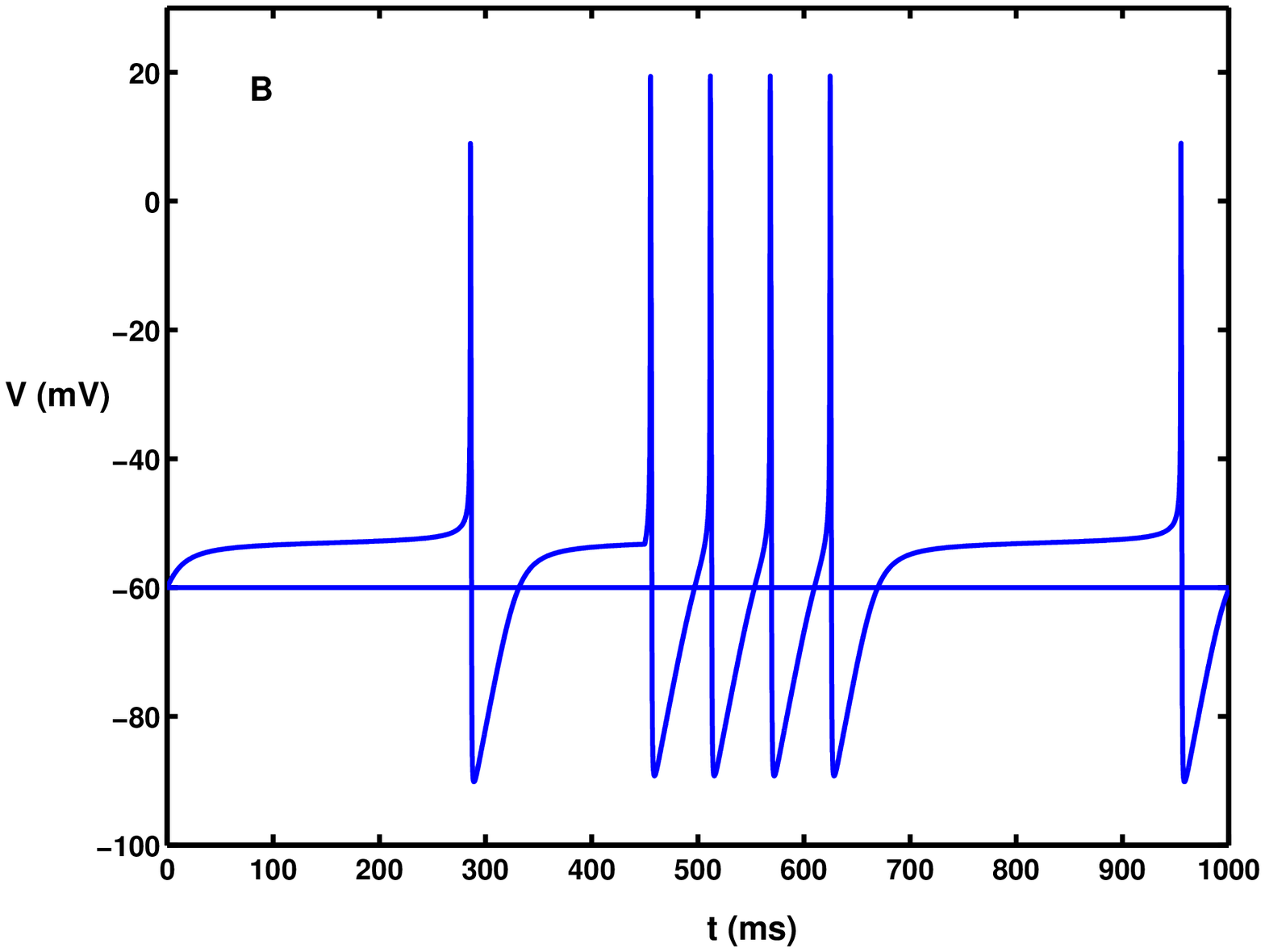,width=2.7in}
\end{center}
\begin{center}
\centerline\leavevmode\epsfig{file=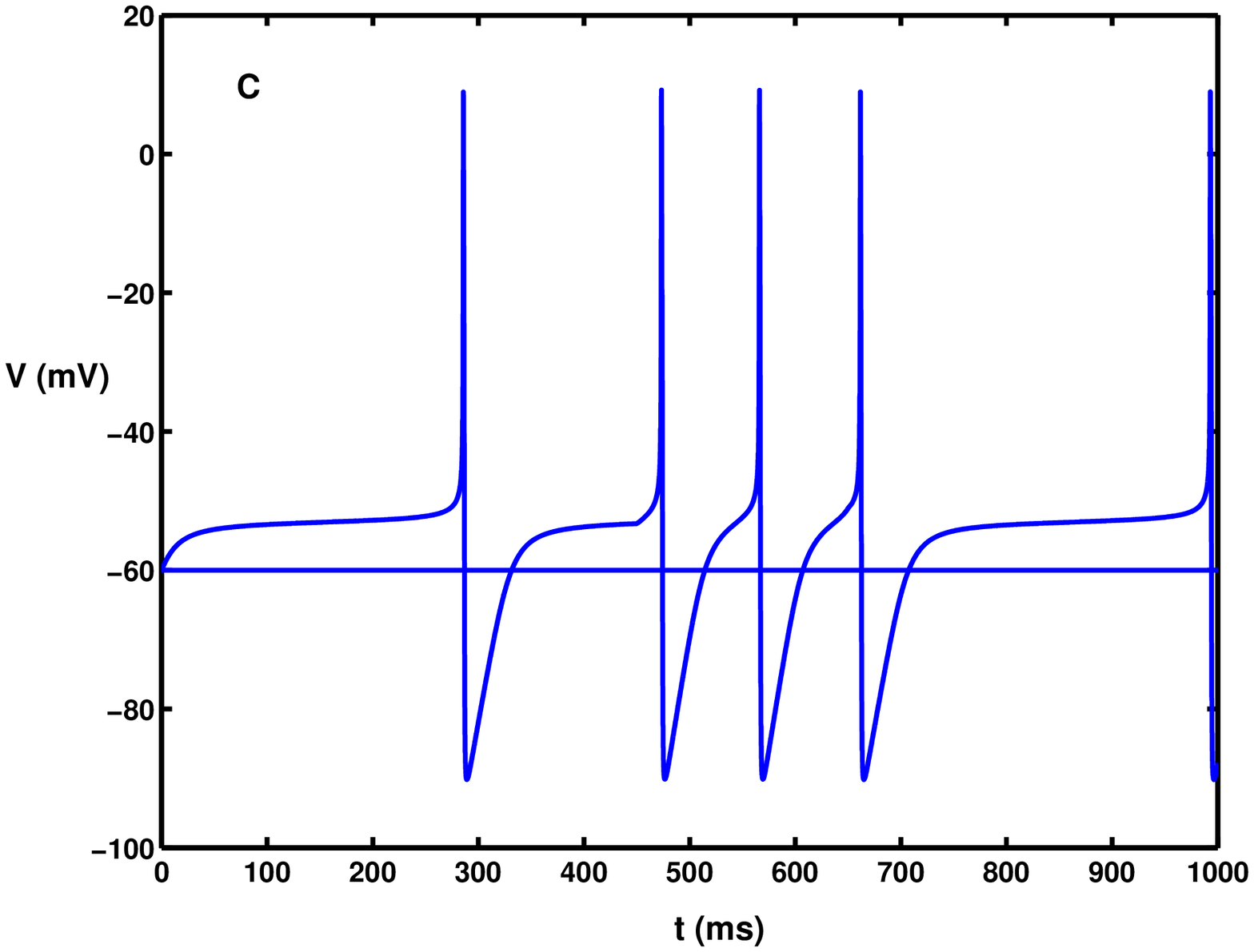,width=2.7in}\epsfig{file=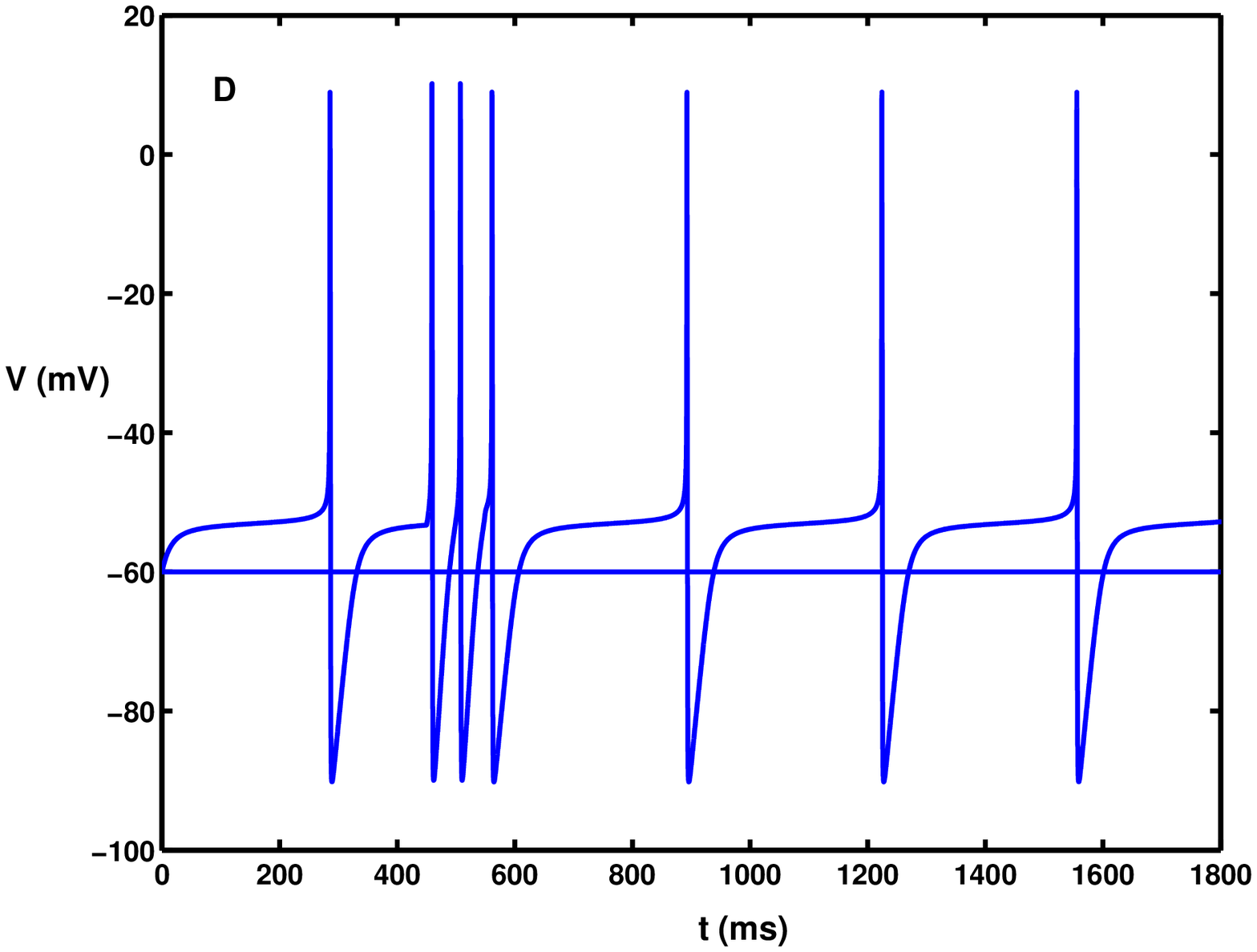,width=2.7in}
\end{center}
\caption{Induction of bursting-type spiking by means of brief changes in excitatory or inhibitory parameters. Parameters are as in set 1 except
with variations in a single parameter as described. 
{\bf A}. Increasing the maximal excitatory conductance $g_{e,max}$
from the standard value of 2 to 2.5 for $450 < t < 650$ gives rise to a burst of three spikes, followed by a return to pacemaker activity. 
{\bf B}. Reducing the maximal inhibitory conductance $g_{i,max}$
from the standard value of 0.5 to 0.3 for the same time period as
in A produces a burst of 4 spikes with a subsequent return to pacemaker-type spiking.  {\bf C}. A 10\% increase in the depolarizing
current for the same time-period as in A and B gives rise to 
rapid spiking followed by the resumption of regular pacemaking activity.
{\bf D}. As in C but now a 50\% increase in depolarizing current applied for the shorter time period $450 < t < 550$ gives rise to more rapid  bursting-type spiking followed regular pacemaking activity.} 
\label{fig:wedge}
\end{figure}

\newpage

\section{Discussion}

Principal brainstem neurons, particularly serotonergic cells of the dorsal raphe nucleus and noradrenergic cells of the locus coeruleus are
of great importance in the functioning of many neuronal populations
throughout cortical and subcortical structures. Of note is the modulatory
role the firing of neurons in these brainstem nuclei have on neurons of the prefrontal cortex, including the 
orbitofrontal cortex, and hippocampus.  These latter structures have
been strongly implicated in various pathologies, including depression,
and OCD. Lanfumey et al. (2008) contains
a comprehensive summary of many of the biological processes which are
influenced by serotonin including those originating
from  serotonergic neurons of the DRN. 
Modeling networks involving both serotonergic and noradrenergic afferents
requires plausible models for the spiking activity of the principal
SE and NA cells. Whereas detailed models of such activity are
now available, their application to many thousands of cells has the
disadvantage of leading to very large computation time and large memory requirements, so that  the simplified models described in the present article
may provide useful approximating  components for such complex computing tasks. 
Accurate models of brainstem neurons (see below) involve
many component currents and very large numbers of parameters,
several of which are possibly uncertain. Hence
realistic simplified mathematical models of brainstem neurons, 
beyond that provided by extremely simplified models such as
the leaky integrate and fire (or Lapicque) model (Tuckwell, 1988), are useful 
in order to investigate approximately the responses of these cells to their complex array of synaptic and other input and to construct and analyze complex networks
involving these cells and those in other centers such as hippocampus,
frontal cortex and hypothalamus. 

The simplified model we have used is a modified Hodgkin-Huxley model
with currents relabeled generically.  With certain parameter sets
such a model predicts spike trajectories which are similar to those
of brainstem neurons in pacemaker mode. The sensitivity of solutions to 
changes in parameters was investigated in detail and very small changes in some parameters were found to produce dramatic changes
in the spiking behavior. Details of such features  as plateaux and 
AHPs could be adjusted to match experimental results with 
judicious choices of voltage parameters. A preliminary investigation 
was able to convert regular pacemaking to bursting behavior in accordance with experimental findings on the application of kynurenic acid to DA and NA neurons.  We end with a brief discussion of previous
computational modeling of DA, NA and SE  brainstem neurons 

\subsection{LC NA neurons}
Thus far there have been several mathematical models
of locus coeruleus neurons per se, which include a few ionic channels
(Putnam et al., 2014; Contreras et al., 2015)  or many ionic channels
including the usual sodium and potassium, high and low threshold
calcium currents, transient potassium $I_A$, persistent sodium,
leak and hyperpolarization activated cation current $I_h$ (De Carvalho et al., 2000;  Alvarez et al. 2002; 
Carter et al., 2012). Noteworthy is the omission of $I_A$ in the 
model of Alvarez et al. (2002) and its inclusion in 
De Carvalho et al. (2000) and Carter et al. (2012).  
Also, a persistent sodium current is included in De Carvalho et al. (2000) and Alvarez et al. (2002) but not in Carter et al. (2012). 

Despite such uncertainties in the mechanisms involved in
pacemaker activity in LC neurons,  some of these works have included 
synaptic input and gap-junction inputs from neighboring LC neurons.
The pioneering article of De Carvalho et al. (2000) addressed the mechanisms of morphine addiction and included 
several biochemical reactions involving cAMP, 
 $\mu$-opioid receptors,  morphine, G-protein, AC, CREB and Fos. 
Tuckwell (2017) contains a summary of previous LC modeling as well as a review of LC neuron anatomy and physiology.  
Brown et al. (2004) did employ Rose-Hindmarsh model neurons to study a network of LC neurons but 
 there have not appeared any
plausible simplified models of these cells per se. Thus the 
two-component model considered in this article provides a good starting
point for investigating, for example, the effects of synaptic input on LC firing which will be performed in future articles.  
\subsection{DRN SE neurons}
For serotonergic neurons of the dorsal raphe there has been only
one detailed model as described in the introduction (Tuckwell and
Penington, 2014). Some authors have addressed quantitatively serotonin release
and included the effects of antidepressants but 
without an explicit model for SE cell spiking (Geldof et al., 2008).
Wong-Lin et al. (2012) use a quadratic integrate and fire model for
spiking DRN SE neurons in a network of such cells along with inhibitory 
neurons. The model has a reset mechanism after spikes which 
is artificial. In a related work, Cano-Colino et al. (2013, 2014) have
modeled the influence of serotonin on networks of excitatory
and inhibitory cells in spatial working memory. More recently,
in a similar vein, Maia and Cano-Colino (2015) have made an interesting
study of serotonergic modulation of the strength of attractors
in orbitofrontal cortex and related this to the occurrence of
OCD. 

\subsection{Midbrain DA neurons}
Midbrain dopaminergic neuons have been implicated, inter alia,  in reward pathways and in several pathological conditions including 
Parkinson's disease (Kalia and Lang, 2015), psychiatric conditions  such as schizophrenia (Weinstein et al., 2017) and addiction to drugs and natural rewards (Nestler, 2004). 
There have been several comprehensive computational modeling studies of these neurons, many of which were carried out by Canavier and coworkers since 1999. These include
the model of Amini et al. (1999) with four calcium currents, two potassium currents, a hyperpolarization activated current and 
various pump currents which was used to explore the nature and role of a slow oscillatory potential which was observed
when spiking was blocked by tetrodotoxin. Compartmental models were considered subsequently by Komendantov et al. (2004) and Kuznetsova et al. (2010), with soma and two
dendritic compartments. In the second of these works it was found that spontaneous firing rates were determined mainly 
by an L-type calcium current and the A-type potassium current.  
In a more recent experimental and computational study using a compartmental model, Tucker et al. (2012) found that pacemaker frequency was dependent on the density of Na$_v$ channels in the soma and their spatial distribution over the soma-dendritic regions.

\section{Acknowledgements}
This research was partially supported by
a Mathematical Biosciences Institute postdoctoral fellowship to YZ
 from the National Science Foundation under Agreement No. 0931642.

%
%

\section{References}

\nh Aghajanian, G.K., Vandermaelen, C.P., 1982. 
Intracellular recordings from serotonergic dorsal raphe neurons: pacemaker
potentials and the effect of LSD. Brain Res. 238,  463-469.

\nh Aghajanian, G.K., Vandermaelen, C.P., Andrade, R., 1983. 
Intracellular studies on the role of calcium in regulating the activity and
reactivity of locus coeruleus neurons in vivo. Brain Res. 273, 237-243.

\nh Alreja, M.,  Aghajanian, G.K., 1991.  Pacemaker activity of locus
 coeruleus neurons: whole-cell recordings in
brain slices show dependence on cAMP and protein kinase A.
Brain Res. 556,  339-343.

\nh Alvarez, V.A., Chow, C.C., Van Bockstaele, E.J., Williams, J.T., 2002.
Frequency-dependent synchrony in locus ceruleus:
role of electrotonic coupling. PNAS 99, 4032-4036.

\nh Amini, B.,  Clark, J.W. Jr, Canavier, C.C., 1999. Calcium dynamics 
underlying pacemaker-like and burst
firing oscillations in midbrain dopaminergic neurons:
A computational study. J. Neurophysiol. 82, 
2249-2261.

\nh Andrade. R., Aghajanian, G.K., 1984. 
Locus coeruleus activity in vitro: intrinsic
regulation by a calcium-dependent potassium
conductance but not $\alpha_2$-adrenoceptors
J. Neurosci 4, .161-170. 

\nh Aston-Jones, G., Bloom, F.E., 1981. 
Activity of norepinephrine-containing locus coeruleus
neurons in behaving rats anticipates fluctuations in
the sleep-waking cycle. J. Neurosci. 1, 876-886. 

\nh Bayliss, D.A., Li, Y.-W., Talley, E.M., 1997a. Effects of serotonin on caudal raphe
neurons: activation of an inwardly rectifying potassium conductance. J. Neurophysiol.
77, 1349–1361. 

\nh  Belluzzi, O., Sacchi, O.,1991. A five-conductance model of the action potential
in the rat sympathetic neurone. Prog. Biophys. Molec. Biol. 55: 1-30

\nh Berridge, C.W., Waterhouse, B.D., 2003. 
The locus coeruleus–noradrenergic system: modulation of behavioral
state and state-dependent cognitive processes. Brain Res. Rev. 42, 33-84.

\nh Blythe, S.N., Wokosin, D., Atherton, J.F., Bevan, M.D., 2009. Cellular mechanisms underlying burst firing in substantia nigra
dopamine neurons.  J. Neurosci. 29, 15531-15541.

\nh Brown, E., Moehlis, J., Holmes, P. et al., 2004.
The influence of spike rate and stimulus duration
on noradrenergic neurons. J. Comp. Neurosci. 17, 13-29. 

\nh Bush, P.C., Sejnowski, T.J., 1993.  
Reduced compartmental models of neocortical pyramidal cells.
Journal of Neuroscience Methods 46,159-166.

\nh Cano-Colino, M., Almeida, R., Compte, A., 2013. 
Serotonergic modulation of spatial working memory: predictions from a computational network model. Frontiers in Integrative Neuroscience 7, 71. 

\nh Cano-Colino, M., Almeida, R., Gomez-Cabrero, D. et al., 2014. 
Serotonin regulates performance nonmonotonically in a spatial working
memory network. Cerebral Cortex 24, 2449-2463. 

\nh Carrasco, G.A., Van de Kar, L.D., 2003. 
Neuroendocrine pharmacology of stress. 
Eur. J. Pharm. 463, 235-272.

\nh Carter, M.E., Brill, J., Bonnavion, P. et al., 2012.
Mechanism for hypocretin-mediated
sleep-to-wake transitions. PNAS 109, E2635–E2644.

\nh Cherugi, K., Charl\'ety, P.J., Akaoka, H. et al., 1993. Tonic activation of NMDA receptors causes spontaneous burst discharge of rat midbrain dopamine neurons in vivo. Eur. J. Neurosci. 5,137-144.

\nh Contreras, S., Quintero, M., Putnam, R., Santin, J., Hartzler, L. and Cordovez, J., 2015. A Computational Model of Temperature-Dependent Intracellular pH Regulation. The FASEB Journal, 29(1 Supplement), pp.860-10

\nh De Carvalho, L.A.V., De Azevedo, L.O., 2000.
A model for the cellular mechanisms
of morphine tolerance and dependence. 
Math. Comp.  Mod. 32,  933-953.

\nh De Oliveira, R.B., Howlett, M.C.H., Gravina, F.S. et al., 2010.
Pacemaker currents in mouse locus coeruleus neurons. Neurosci 170, 166-177. 

\nh De Oliveira, R.B., Gravina, F.S.,  Lim, R. et al., 2011.
Developmental changes in pacemaker currents in mouse locus
coeruleus neurons. Brain Res. 1425, 27-36. 

\nh Engberg, G., Kling-Peterson, T., Nissbrandt, H., 1993.
GABA$_B$-receptor activation alters the
firing pattern of dopamine neurons in the
rat substantia nigra. Synapse 15, 229-238.

\nh Fitzhugh, R.,1961. Impulses and physiological states in theoretical models of nerve membrane. Biophys. J. 1, 445-466.

\nh Flower, G., Wong-Lin, K., 2014. 
Reduced computational models of serotonin
synthesis, release and reuptake.
IEEE Trans. Biomed. Eng. 61, 1054-1061.

\nh Foote, S.L., Aston-Jones, G., Bloom, F.E., 1980. 
Impulse activity of locus coeruleus neurons in awake rats and
monkeys is a function of sensory stimulation and arousal. 
Proc. Nati. Acad. Sci. USA 77, 3033-3037. 

\nh Geldof, M., Freijer,  J.I., Peletier, L.A. et al., 2008.
Mechanistic model for the acute effect of fluvoxamine on
5-HT and 5-HIAA concentrations in rat frontal cortex. 
Eur. J. Pharm. Sci. 33, 217-219.

\nh German, D.C. and Manaye, K.F., 1993. Midbrain dopaminergic neurons (nuclei A8, A9, and A10): Three-dimensional reconstruction in the rat. J. Comp. Neurol. 331, 297-309.

\nh Grace, A. A., Bunney, B.S., 1983.  Intracellular and extracellular
electrophysiology of nigral dopaminergic neurons. I. Identification and
characterization. Neurosci. 10: 301-315.

\nh Grace, A.,  Bunney, B., 1984.  The control of firing pattern in nigral dopamine neurons: burst firing. J. Neurosci. 4,  2877-2890. 

\nh Grace, A.A., Onn, S-P., 1989. Morphology and electrophysiological properties of immunocytochemically identified rat dopamine
neurons recorded in vitro. J. Neurosci. 9, 3463-3481.

\nh Grace, AA., Bunney, B.S., 1995. 
Electrophysiological Properties of Midbrain Dopamine Neurons.
Neuropsychopharmacology - 4th Generation of Progress. Eds. Bloom, F.E., Kupfer, D.J. Raven Press, New York.

\nh Grenhoff, J., Tung, C-S., Svensson, T.H., 1988. 
The excitatory amino acid antagonist kynurenate
induces pacemaker-like firing of dopamine neurons
in rat ventral tegmental area in vivo. Acta Physiol. Scand. 134, 
567-568. 

\nh H\'aj\'os, M.,  Sharp, T., Newberry, N.R., 1996. 
Intracellular recordings from burst-firing presumed serotonergic neurones
the rat dorsal raphe nucleus in vivo. Brain Res. 737, 308-312.

\nh H\'aj\'os,  M.,  Allers, K.A., Jennings, K. et al., 2007. 
Neurochemical identification of stereotypic burst-firing
neurons in the rat dorsal raphe nucleus using juxtacellular
labelling methods. Eur. J. Neurosci. 25, 119-126.

\nh Harris, N.C., Webb, C., Greenfield, S.A., 1989. A possible pacemaker mechanism in pars compacta neurons of the guinea-pig substantia nigra revealed by various ion channel blocking agents. Neurosci. 31, 355-362.

\nh Hodgkin, A.L., 1948.  The local changes associated with repetitive action in a
non-medullated axon. J. Physiol. 107, 165-181.

 \nh  Hodgkin, A.L., Huxley, A.F., 1952.  A quantitative description of membrane
current and its application to conduction
and excitation in nerve.  J. Physiol. 117, 500-544.

\nh Jedema, H.P., Grace, A.A., 2004. 
Corticotropin-releasing hormone directly activates
noradrenergic neurons of the locus ceruleus recorded
in vitro. J. Neurosci. 24, 9703-9713.

%

\nh Ishimatsu, M., Williams, J.T., 1996. Synchronous activity in locus coeruleus results from dendritic
interactions in pericoerulear regions. J. Neurosci. 16, 5196-5204.

\nh Jacobs, B.L., Azmitia, E.C., 1992. Structure and function of the brain
 serotonin system.
   Physiol. Rev. 72, 165-229.

\nh Jo\"els, M, Karst, H., Krugers, H.J., Lucassen, P.J., 2007.
Chronic stress: Implications for neuronal morphology, function
and neurogenesis. Front. Neuroendocrin. 28, 72-96. 

\nh Joshi, A., Youssofzadeh, V., Vemana, V., McGinnity, T.M., Prasad, G., Wong-Lin, K., 2017. An integrated modelling framework for neural circuits with multiple neuromodulators. Journal of The Royal Society Interface 14, 20160902.

\nh Kalia, L.V., Lang, A.E., 2015.  Parkinson’s disease. Lancet 386, 896-912.

\nh Khaliq, Z.M., Bean, B.P., 2008. 
Dynamic, nonlinear feedback regulation of slow
pacemaking by A-type potassium current in ventral
tegmental area neurons. J. Neurosci. 28, 10905-10917.

\nh  Kirby, L.G.,  Pernar, L.,  Valentino, R.J. et al., 2003. Distinguishing characteristics of serotonin and nonserotonin-
containing cells in the dorsal raphe nucleus:
electrophysiological and immunohistochemical studies. Neurosci. 116,  669-683.

\nh Kocsis B, Varga V, Dahan L, Sik A (2006)  
Serotonergic neuron diversity: Identification
of raphe neurons with discharges time-locked
to the hippocampal theta rhythm.
 PNAS 103: 1059-1064.  

\nh \nh  Komendantov, A.O., Komendantova, O.G., Johnson, S.W., Canavier, C.C., 2004.
A modeling study suggests complementary roles for GABA$_A$ and NMDA
receptors and the SK channel in regulating the firing pattern in midbrain
dopamine neurons. J. Neurophysiol. 91, 346-357.

\nh Korf, J., Bunney, B.S., Aghajanian, G.K., 1974.
 Noradrenergic neurons: morphine inhibition
of spontaneous activity. Eur.J. Pharmacol. 25, 165-169.

\nh Kubista, H., Boehm, S., 2006. Molecular mechanisms underlying the modulation of exocytotic
noradrenaline release via presynaptic receptors. 
Pharmacol. Therap. 112, 213-242.

\nh Kuznetsova, A.Y., Huertas, M.A., Kuznetsov, A.S., Paladini, C.A.,  Canavier, C.C., 2010. Regulation of firing frequency in a computational model of a midbrain dopaminergic neuron. Journal of computational neuroscience 28, 389-403.

\nh Lanfumey ,L., Mongeau, R., Cohen-Salmon, C., Hamon, M.,  2008.
Corticosteroid-serotonin interactions in the neurobiological mechanisms
of stress-related disorders. Neurosci. Biobehav. Rev. 32, 1174-1184.

\nh Levine, E.S.,  Jacobs, B.L., 1992.
Neurochemical afferents controlling the activity of serotonergic
neurons in the dorsal raphe nucleus: microiontophoretic
studies in the awake cat.  J. Neurosci. 12, 4037-4044. 

\nh Li, Y-Q., Li, H., Kaneko, T., Mizuno, N., 2001.
Morphological features and electrophysiological properties of
serotonergic and non-serotonergic projection neurons in the dorsal
raphe nucleus
An intracellular recording and labeling study in rat brain slices. 
Brain Res. 900, 110-118.

\nh  Li, Y-W., Bayliss, D.A., 1998. Electrophysiological properties, synaptic transmission
and neuromodulation in serotonergic caudal raphe neurons. 
Clin. Exp. Pharm. Physiol.  25, 468-473.

\nh Lopez. J.F., Akil, H., Watson, S.J., 1999.
Neural circuits mediating stress.  Biol. Psychiatry 46:1461-1471.

\nh Lowry, C.A., Evans, A.K., Gasser, P.J. et al., 2008.
 Topographic organization and chemoarchitecture of
the dorsal raphe nucleus and the median raphe nucleus. In:
 Serotonin
and sleep: molecular, functional and clinical aspects,  p 25-68, 
Monti, J.M. et al., Eds.
Basel: Birkhauser Verlag AG.

\nh Lowry, C.A., Rodda, J.E., Lightman, S.L., Ingram, C.D., 2000. Corticotropin-
releasing factor increases in vitro firing rates of serotonergic
neurons in the rat dorsal raphe nucleus: evidence for activation of a
topographically organized mesolimbocortical serotonergic system.
J. Neurosci. 20, 7728-7736.

\nh Luppi. P-H., Clement, O., Sapin, E. et al., 2012. 
Brainstem mechanisms of paradoxical (REM)
sleep generation. Eur. J. Physiol. 463:43-52.

\nh Maejima, T., Masseck, O.A., Mark, M.D., Herlitze, S., 2013. 
Modulation of firing and synaptic transmission of serotonergic neurons by intrinsic G protein-coupled receptors and ion channels.
Frontiers in Integrative Neuroscience 7, 40.

\nh Mahar, I., Bambico, F.R., Mechawar, N., Nobrega J.N., 2014.
Stress, serotonin, and hippocampal neurogenesis in relation to
depression and antidepressant effects. Neurosci. Biobehav. Rev.
38, 173-192.

\nh Maia, T.V., Cano-Colino, M., 2015.   The role of serotonin in orbitofrontal
function and obsessive-compulsive
disorder. Clinical Psychological Science 3, 460-482, and Supplemental-data. 
%

\nh Milnar, B., Montalbano, A., Piszczek et al., 2016. 
Firing properties of genetically
identified dorsal raphe serotonergic
neurons in brain slices. Front. Cell. Neurosci. 10, 195.

\nh Nagumo, J. S., Arimoto, S.,  Yoshizawa, S., 1962.  An active pulse transmission line
simulating nerve axon. Proc. I.R.E. 50, 2061-2070.

\nh Nair-Roberts, R.G., Chatelain-Badie, S.D., Benson, E., White-Cooper, H., Bolam, J.P.,  Ungless, M.A., 2008. Stereological estimates of dopaminergic, GABAergic and glutamatergic neurons in the ventral tegmental area, substantia nigra and retrorubral field in the rat. Neurosci. 152, , 1024-1031.

\nh Nestler, E.J., 2004. Molecular mechanisms of drug addiction.
Neuropharmacol. 47, 24-32.

\nh Nestler, E. J., Alreja, M., Aghajanian, G.K., 1999.
Molecular control of locus
coeruleus neurotransmission. 
Biol. Psychiatry 46, 1131-1139.

\nh Ohliger-Frerking, P.,  Horwitz, B.A., Horowitz,  J.M., 2003. 
Serotonergic dorsal raphe neurons from obese zucker
rats are hyperexcitable. Neurosci. 120, 627-634.

\nh Overton, P., Clark., D.,1992.  Iontophoretically administered drugs acting at the N-methyl-D-aspartate receptor modulate burst firing in A9 dopamine neurons in the rat. Synapse 10,131-140.

\nh Pan, W.J., Osmanovi\'c, S.S., Shefner, S.A., 1994.
Adenosine decreases action potential duration by modulation of
A-current in rat locus coeruleus neurons.
J. Neurosci. 14, 1114-1122. 

\nh Penington, N.J., Kelly, J.S., Fox, A.P., 1991. A study of the mechanism of Ca2+ current
inhibition produced by serotonin in rat dorsal raphe neurons. J. Neurosci. I7, 3594-3609.

\nh Putnam, R., Quintero, M., Santin, J. et al., 2014. 
Computational modeling of the effects of temperature on chemosensitive locus coeruleus neurons from bullfrogs. Faseb J. 28, Supp. 1128.3. (Abstract only). 

\nh Putzier, I., Kullmann, P.H.M., Horn, J.P., Levitan, E.S., 2009. 
Ca$_v$1.3 channel voltage dependence, not Ca$^{2+}$ selectivity,
drives pacemaker activity and amplifies bursts in nigral
dopamine neurons. J. Neurosci. 29, 15414-15419.

\nh Rall, W., 1962. Theory of physiological properties of dendrites.
Ann. NY Acad. Sci. 96: 1071-1092.

\nh Ramirez, J-M., Koch, H., Garcia, A.J. III, et al., 2011.
The role of spiking and bursting pacemakers
in the neuronal control of breathing. J Biol. Phys. 37, 241-261.


\nh Sanchez-Padilla, J., Guzman, J.N., Ilijic, E. et al., 2014.
Mitochondrial oxidant stress in locus coeruleus is regulated by activity and nitric oxide synthase. Nat. Neurosci. 17, 832-842. 

\nh Svensson, T.H., Engberg, C-S., Tung, C-S., Grenhoff, J., 1989.
 Pacemaker-like firing of noradrenergic locus
coeruleus neurons in vivo induced by the
excitatory amino acid antagonist kynurenate
in the rat. Acta Physiol. Scand. 135, 421-422.

\nh Sugiyama, D., Hur, S.W., Pickering, A:E. et al. 2012.
In vivo patch-clamp recording from locus coeruleus
neurones in the rat brainstem.
J Physiol. 590, 2225-2231.

\nh Swanson, L.W., 1976. The locus coeruleus: a cytoarchitectonic, golgi and
immunohistochemical study in the albino rat. Brain Res. 110, 39-56.



\nh Tucker, K.R., Huertas, M.A., Horn, J.P., Canavier, C.C., Levitan, E.S., 2012. Pacemaker rate and depolarization block in nigral dopamine neurons: a somatic sodium channel balancing act. J. Neurosci 32, 14519-14531.

\nh Tuckwell, H.C., 1988. Introduction to Theoretical Neurobiology. 
 Cambridge University Press, Cambridge UK. 

\nh Tuckwell, H.C., 2013. 
Biophysical properties and computational modeling of calcium spikes in
serotonergic neurons of the dorsal raphe nucleus.
BioSystems 112, 204-213. 

\nh Tuckwell, H.C., 2017. Computational modeling of spike generation in locus coeruleus noradrenergic neurons. Preprint. 

\nh Tuckwell, H.C., Ditlevsen, S., 2016. 
The space-clamped Hodgkin-Huxley system with random
synaptic input: inhibition of spiking by weak noise
and analysis with moment equations. Neural Computation 28, 2129-2161.

\nh Tuckwell, H.C., Jost, J., GutKin, B.S., 2009. Inhibition 
and modulation of rhythmic neuronal spiking by noise. Phys. Rev. E. 80,
031907. 

\nh Tuckwell, H.C., Penington, N.J., 2014. Computational modeling of spike
generation in serotonergic neurons of the dorsal raphe nucleus.
Prog. Neurobiol. 118, 59-101. 

\nh Valentino, R.J., Foote, S.L., Page, M.E., 1993.
The locus coeruleus as a site for
integrating corticotropin-releasing
factor and noradrenergic mediation
of stress responses. Ann. NY.Acad. Sci 697, 173-188.

\nh Vandermaelen, C.P., Aghajanian, G.K., 1983. Electrophysiological and pharmacological
characterization of serotonergic dorsal raphe neurons recorded extracellularly
and intracellularly in rat brain slices. Brain Res. 289, 109–119.

\nh Vasudeva, R.K., Waterhouse, B.D., 2014. 
Cellular profile of the dorsal raphe lateral wing sub-region:
relationship to the lateral dorsal tegmental nucleus. J. Chem. Neuroanat. 57-58,
15-23.

\nh Vertes, R.P., Crane, A.M., 1997. 
Distribution, quantification,
and morphological characteristics
of serotonin-immunoreactive cells
of the supralemniscal nucleus (b9)
and pontomesencephalic reticular
formation in the rat. J. Comp. Neurol. 378, 411-424. 

\nh Walsh, J.B., Tuckwell, H.C., 1985.  Determination of the electrical
potential over dendritic trees by mapping onto a nerve cylinder.
J. Theor. Neurobiol. 4, 27-46. 

\nh Weinstein, J.J., Chohan, M.O., Slifstein, M., Kegeles, L.S., Moore, H., Abi-Dargham, A., 2017. Pathway-specific dopamine abnormalities in schizophrenia. Biological psychiatry, 81(1), pp.31-42.

\nh Williams, J.T., Egan, T.M., North, R.A., 1982. Enkephalin opens 
potassium channels on mammalian central neurons. Nature 299, 74-77. 


\nh Williams, J.T., North, R.A., Shefner S.A. et al., 1984.
Membrane properties of rat locus coeruleus
neurones. Neuroscience 13, 137-156.

\nh Wong-Lin, K-F., Joshi, A., Prasad, G., McGinnity, T.M., 2012. 
Network properties of a computational model of the dorsal raphe nucleus. Neural Netw. 32, 15-25.
%

%

\end{document}